\documentclass[twoside,12pt]{article}
\usepackage{epsfig}
\usepackage{graphicx}
\usepackage{amssymb}
\newcommand{\be}{\begin{equation}}
\newcommand{\ee}{\end{equation}}
\newcommand{\beq}{\begin{eqnarray}}
\newcommand{\eeq}{\end{eqnarray}}

\begin{document}

\title{Propagation of vector-meson spectral-functions in a BUU type transport model:\\
Application to di-electron production}
\author{{\sc H.W. Barz, B. K\"ampfer, Gy. Wolf, M. Z\'et\'enyi}\\[3mm]
$^1$Forschungszentrum Dresden-Rossendorf, Institut f\"ur Strahlenphysik,\\
PF 510119, 01314 Dresden, Germany\\[1mm]
and\\[1mm]
$^2$KFKI RMKI, H-1525 Budapest, POB 49, Hungary} 

\maketitle

\begin{abstract}
The time evolution of vector-meson spectral-functions is studied
within a kinetic theory approach.
We implement this formalism in a BUU type transport model.
Applications focus on $\rho$ and $\omega$ mesons being
important pieces for the interpretation of
the di-electron invariant mass spectrum measured by the HADES collaboration for
the reaction \mbox{C + C} at 2 AGeV bombarding energy.
Since the evolution of the spectral functions is driven by the local
density, the in-medium modifications are tiny for small collision systems
within this approach.
\end{abstract}

\section{Introduction}

Di-electrons serve as direct probes of dense and hot nuclear matter stages 
during the course of heavy-ion collisions \cite{RW,Itzak}. 
The superposition of various sources,
however, requires a deconvolution of the spectra by means of models.
Of essential interest are the contributions of the light vector mesons $\rho$ and $\omega$. 
The spectral functions of both mesons are expected to be modified in a strongly interacting
environment in accordance with chiral dynamics, QCD sum rules etc.\ 
\cite{RW,Hatsuda_Lee,Brown_Rho,MML}. 
After the first pioneering experiments with the Dilepton Spectrometer DLS \cite{DLS}
now improved measurements with the High-Acceptance Di-Electron Spectrometer 
HADES \cite{HADES,HADES_PRL,HADES_1GeV} 
start to explore systematically 
the baryon-dense region accessible in fixed-target heavy-ion experiments at beam
energies in the 1 - 2 AGeV region.
The invariant mass spectra of di-electrons for the reaction C + C
at 1 and 2 AGeV are now available \cite{DLS,HADES_PRL,HADES_1GeV} allowing us to
hunt for interesting many-body effects.

There are several approaches for describing the emission of real and virtual 
photons off excited nuclear matter:

(i) A piece of matter in thermal equilibrium at temperature $T$ emits $e^+ e^-$
pairs with total momentum $q$ at a rate 
$dN/d^4xd^4q\,=\, \alpha^2/(M^2\pi^3) f_B  Im \Pi_{em}$,
where $f_B$ is the bosonic thermal distribution function 
and $\Pi_{em}$ denotes the 
electromagnetic current-current correlator, 
$\Pi_{em} \,=\, -i\int d^4x e^{iqx} \langle\langle {\cal T} j^\mu(x)j_\mu(0) \rangle\rangle$, 
the imaginary part of which determines directly
the thermal emission rate. Here, $\alpha$ stands for the electromagnetic fine 
structure constant, $M$ means the invariant mass of the di-electron, 
${\cal T}$ is the operator
for time ordering  which acts on the electromagnetic current operator $j^\mu$. 
This rate may be combined
with  a global dynamic model which provides us the space averaged quantities as 
a function of time \cite{Rapp} or even adopting also a time average 
\cite{Gallmeister}.

(ii) Some sophistication can be achieved by employing a detailed model for the space-time
evolution of baryon density and temperature. e.g., as delivered by hydrodynamics.
One may also extract from transport models such parameters, where, however, local 
off-equilibrium and/or anisotropic momentum distributions hamper a reliable definition
of density and temperature. Nevertheless, once the rate is given one has a
very concise approach, as realized,  e.g., in \cite{Kapusta_hydro}.
A similar approach has been presented in \cite{Wambach_Cassing}.

(iii) Microscopic or kinetic transport models do not require isotropic
momentum distributions or local equilibrium. 
Once general principles are implemented, transport models also provide
a detailed treatment of the emission of electromagnetic radiation in heavy-ion collisions.

Our approach belongs to item (iii). The time evolution of 
single particle distribution functions
of various hadrons are evaluated within the framework of a kinetic theory. 
We focus on the vector mesons $\rho$ and $\omega$. The $\rho$ meson is
already a broad resonance in vacuum, while the $\omega$ meson may acquire a noticeable
width in nuclear matter \cite{broadening_omega}. Therefore, we are forced to treat 
dynamically  these resonances and their decays into di-electrons. 
Resorting to consider only pole-mass
dynamics is clearly insufficient in a microscopic approach \cite{Knoll}. Instead, one
has to propagate properly the spectral functions of the $\rho$ and $\omega$ mesons.
This is the main goal of our paper. We consider our work as being on an explorative
level, not yet as a firm and deep theoretically founded prescription of dealing with 
di-electron emission from excitations with quantum numbers 
of $\rho$ and $\omega$ mesons off excited nuclear matter.

Our paper is organized as follows. Essential features of 
our transport model are outlined in section 2. In Subsection 2.1 we describe how the
mean field potentials enter in the relativistic transport equation.
Subsection 2.2 introduces the dynamics of broad resonances and its implementation
in the test-particle method. The crucial quantities for the spectral functions
are the self-energies dealt with in subsection 2.3. Subsections 2.4 and 2.5 are 
devoted to particle production and di-electron emission, respectively. Numerical 
results of our simulations and a tentative comparison with published HADES data 
are presented in sections 3 (2 AGeV) and 4 (1 AGeV).
Discussion and summary can be found in section 5.

The present analysis supersedes \cite{our_first_attempt}.
Further analyses of HADES data have been performed in
\cite{Brat_Cass,Aichelin,Tuebingen,Bleicher}.

\section{Treatment of heavy-ion collisions}

\subsection{The standard BUU treatment}

The employed BRoBUU computer code for heavy-ion collisions
developed by a Budapest-Rossendorf cooperation
solves a set of coupled Boltzmann-\"Uhling-Uhlenbeck (BUU) equations in the
quasi-particle \mbox{limit \cite{wolf93}} 
\begin{eqnarray}
\label{BUU}
\frac{\partial F_i}{\partial t} + \frac{\partial H}{\partial {\bf p}}
\frac{\partial F_i}{\partial {\bf x}} -  \frac{\partial H}{\partial {\bf x}}
\frac{\partial F_i}{\partial {\bf p}}  = \sum_j{\cal C}_{ij}, \quad 
H = \sqrt{(m_i+U({\bf p},{\bf x}))^2 + {\bf p}^2}
\end{eqnarray}
for the one-body distribution functions $F_i({\bf x},{\bf p},t)$ of 
the various hadron species $i$,
each with rest mass $m_i$, 
in a momentum and density dependent mean field $U$.
The scalar mean field $U$ is chosen in such a manner that the Hamiltonian $H$
equals $H=\sqrt{m_i^2+{\bf p}^2}+U_i^{nr}$ with a potential $U_i^{nr}$
calculated in the local rest frame as 
\begin{equation}
U_i^{nr} = A \frac{n}{n_0} + B \left( \frac{n}{n_0} \right)^\tau
+ C \frac{2}{n_0} \int \frac{d^3 p'}{(2\pi)^3} 
\frac{F_N(x,p')}{1 + \left( \frac{{\bf p} - {\bf p}'}{\Lambda} \right)^2},
\end{equation}
where the parameters $A$, $B$, $C$, $\tau$, $\Lambda$ define special types of 
potentials, while $n$, $n_0$ and $F_N$ stand for the baryon number density, 
saturation density and nucleon distribution function.
We use the momentum dependent soft potential defined by
$A$=-0.120 GeV, $B$=0.151 GeV, $\tau$=1.23, $C$=-0.0645 GeV, $\Lambda$=2.167 GeV.
The BRoBUU code propagates in the baryon sector the nucleons and
24 $\Delta$ and $N^*$ resonances and additionally 
$\pi,\eta,\sigma,\omega$ and $\rho$ mesons.
Different particle species 
are coupled by the collision integral ${\cal C}_{ij}$ which 
also contains the \"Uhling-Uhlenbeck terms responsible
for Pauli blocking of spin-1/2 hadrons in the collision as well as
particle creation and annihilation processes. 

The set of coupled BUU equations is solved by using the
parallel-ensemble test-particle method \cite{method,Wolf90}, where 
we introduce a number of parallel ensembles.
In each ensemble a test particle represents a real particle
(nucleon, resonance, pion etc.); collisions happen only within
the same ensemble. On the other hand, when calculating
such quantities as densities, Pauli blocking factors etc.\
we average over the ensembles in each time step.   
This method transforms the 
partial differential-integro equations (\ref{BUU})
into a set of ordinary differential equations (looking like equations of motion)
for a number of test particles.
A default version of the code has been applied to strangeness 
dynamics \cite{barznaumann}.

\subsection{Off-shell transport of broad resonances}

Recently theoretical progress has been made in describing the in-medium properties
of particles starting from the Kadanoff-Baym equations \cite{Kadanoff} for
the Green functions of particles. 
Applying first-order gradient expansion after a Wigner transformation 
one arrives at a transport equation for 
the retarded Green function \cite{Cassing-Juchem00,Leupold00}.  
In the medium, particles acquire a self-energy $\Sigma(x,p)$  which depends on position 
and momentum  as well as the local properties of the surrounding medium.
They have a finite life time which is 
described  by  the width $\Gamma$  related to the imaginary part of the self-energy.
Their properties are described by the spectral function 
being the imaginary part of the retarded propagator
${\cal A }(p) = -2 Im G^{ret}(x,p)$. For bosons 
the spectral function is related to the self-energy via
\begin{equation}
\label{spectral}
{\cal A }(p) = 
\frac{\hat{\Gamma}(x,p)}{(E^2 -{\vec p}^2-m_0^2-
Re\Sigma^{ret}(x,p))^2 + \frac{1}{4}\hat{\Gamma}(x,p)^2}\,,
\end{equation}
where the resonance widths $\Gamma$ and $\hat{\Gamma}$ obey 
$\hat{\Gamma}(x,p) = - 2 Im \Sigma^{ret}\approx 2m_0\Gamma$,
and $m_0$ is the vacuum pole mass of the respective particle. 
The spectral function (\ref{spectral}) is normalized as
$\int dp^2 {\cal A } = 2\pi$.
(We omit here the label denoting the particle type for simplicity reasons.)

To solve numerically the Kadanoff-Baym equations one may exploit the above 
test-particle ansatz for a modified retarded Green function  
(see Refs.~\cite{Cassing-Juchem00,Leupold00}). This function can be interpreted as
a product of particle number density multiplied with the spectral
function ${\cal A}$. The spectral function can significantly 
change in the course of the heavy-ion collision process.
Therefore, the standard test-particle method, where the test-particle mass is a constant
of motion, must be extended 
by treating the energy $E = p^0$ of the four-momentum $p$
as an independent variable.

Equations of motion for test particles follow from the transport equation.
We use the relativistic version of the equations which have been derived
in Ref.~\cite{Cassing-Juchem00}:
\begin{eqnarray}
  \label{eq:x}
  \frac{d {\vec x}}{dt} & = &
  \frac{1}{1 - C}
  \frac{1}{2 E}
  \left(
    2 {\vec p} + {\vec \partial}_{p} Re \Sigma^{ret} +
    \frac{m^2 - m_0^2 - Re \Sigma^{ret}}{\hat{\Gamma}}
    {\vec \partial}_{p} \hat{\Gamma}
  \right),\\
  \label{eq:p}
  \frac{d {\vec p}}{d t} & = &
  - \frac{1}{1-C}
  \frac{1}{2 E}
  \left(
    {\vec \partial}_{x} Re \Sigma^{ret}
    + \frac{m^2 - m_0^{2}
      - Re \Sigma^{ret}}{\hat{\Gamma}}
    {\vec \partial}_{x} \hat{\Gamma}
  \right),\\
  \label{eq:e}
  \frac{d E}{d t} & = &
  \frac{1}{1-C}
  \frac{1}{2 E}
  \left(
    \partial_{t} Re \Sigma^{ret}
    + \frac{m^2 - m_0^{2}
      - Re \Sigma^{ret}}{\hat{\Gamma}}
    {\partial}_{t} \hat{\Gamma}
  \right),
\end{eqnarray}
with the renormalization factor reads
\begin{equation}
  \label{eq:C}
  C = \frac{1}{2 E}
  \left(
  {\partial_E}Re \Sigma^{ret}
  + \frac{m_n^2 - m_0^2 - Re \Sigma^{ret}}{\hat{\Gamma}}
  {\partial_E}\hat{\Gamma}
\right).
\end{equation}
In the above, $m = \sqrt{E^2 - {\vec p}^2}$ is the mass of 
an individual 
test-particle number of a given hadron specie.   
The self-energy $\Sigma^{ret}$
is considered to be a function of density $n$,
energy $E$, and momentum $\vec p$; thus the dependence
on time and position comes only from its density dependence.
Partial derivatives with respect to any
of the four variables, $t,{\vec x},E,{\vec p}$,
are understood taking the three other ones fixed. The quantity $1-C$ 
has been introduced to ensure that the test particles 
describe a conserved quantity \cite{Leupold00}.

The change of the test-particle mass $m$ can be more clearly seen combining
Eqs.~(\ref{eq:p}) and (\ref{eq:e}) to
\be
\label{eq:m}
\frac{dm^2}{dt} \,=\, \frac{1}{1-C} \left(\frac{d}{dt} {Re \Sigma^{ret}} +
\frac{m^2 - m_0^2 - Re \Sigma^{ret}}{\hat \Gamma}
{\frac{d}{dt} {\hat \Gamma}} \right)
\ee
with the comoving derivative $d/dt \equiv \partial_t + {\vec p}/E
{\vec \partial}_x$.
This equation means that the square of the particle mass tends
to reach a value shifted by the real part of the self-energy
within a range of the value of ${\hat \Gamma}$. Thus, the
vacuum spectral function is
recovered when the particle leaves the medium.
This ensures the smooth transition from the in-medium behavior
to the vacuum properties.

The equation of motions of the test particles 
have to be supplemented by a collision term
which couples the equations for the different particle species.
It can be shown \cite{Leupold00} that 
this collision term has the same form as in the
standard BUU treatment.

The off-shell transport has been  implemented in simulations
for the propagation of $\rho$ and $\omega$ 
mesons to study the di-electron productions in  $\gamma A$ 
and  $pA$ reactions. Early approaches \cite{Eff_gammaA,Eff_off-shell}
did not automatically provide 
the correct asymptotic behavior of the spectral function and
an auxiliary potential were introduced  to cure this problem.
The above equations of motion, which do not have this deficit,
were applied to di-electron production
in $pA$ collisions in \cite{Bratkovskaya} and for $A A$ collisions
in \cite{Brat_Cass}.

\subsection{Self-energies}

To solve the Eqs.~(\ref{eq:x}-\ref{eq:e}) 
one needs the knowledge of the self-energies.
Here one faces the need to decide 
which effects to take into account in
the expression for the retarded self-energy $\Sigma^{ret}$ in the medium.
That is because the BUU transport equations themselves already contain some 
part of in-medium effects that usually are considered in theoretical models
in local density and local equilibrium approximation \cite{Lutz,rho_models,rho_models1}.
For instance models for in-medium effects of $\rho$ mesons 
usually  take into account the
$N(1520)$--nucleon-hole loop for the self-energy, 
the corresponding vertices are accounted for in BUU via
$\rho$-nucleon scattering and absorption through the $N(1520)$ resonance. 

In our calculations we  employ a simple schematic
form of the self-energy of a vector meson $V$:
\begin{eqnarray} \label{areal}
{\rm Re} \Sigma^{ret}_V & = & 2 m_V \Delta m_V \frac{n}{n_0},\\
\label{aimag}
{\rm Im} \Sigma^{ret}_V & = & m_V (\Gamma^{vac}_V +
\frac{n v \sigma_V}{\sqrt{1-v^2}}).
\end{eqnarray}
Equation~(\ref{areal}) causes a ''mass shift'' 
$\Delta m = \sqrt{m_V^2+Re\Sigma_V^{ret}}-m_V$
characterized by $\Delta m_V$ and being roughly proportionally to the
density $n$ of the surrounding matter.
The imaginary part contains the vacuum width $\Gamma^{vac}_V$ 
the energy dependence of which is described by a form factor \cite{wolf97}.
The second term in Eq.~(\ref{aimag}) results from the collision broadening which
depends on density, relative velocity $v$ and the cross section $\sigma_V$
of the vector meson in matter. This cross section $\sigma_V$ 
is calculated via the Breit-Wigner formula
\begin{equation}
\sigma_V =  \frac{4\pi}{q_{in}^2} \sum\limits_R \frac{2J_R+1}{3(2J_i+1)}
\frac{s \Gamma_{V,R}\Gamma^{tot}_R}{(s-m_R^2)^2 +
s(\Gamma^{tot}_R)^2}
\end{equation}
for forming resonances with masses $m_R$, angular momenta $J_{R}$, partial widths
$\Gamma_{V,R}$, total widths $\Gamma^{tot}_R$ with energy $\sqrt{s}$ and relative
momentum $q_{in}$ in the entrance channel.
In vacuum the baryon density $n$ vanishes and the resulting spectral
function ${\cal A}_{vac}$ is solely determined
by the energy dependent width $\Gamma_V^{vac}$.
We remark that the decay of a test particle is determined
by its vacuum width. The life time of a test particle is, furthermore,
reduced by the absorption during
the two-particle collisions which characterizes 
the total width ${\rm Im}  \Sigma^{ret}_V$.

In our actual numerical implementation we assume  
that the spectral function of the $\rho$ ($\omega$) meson 
vanishes below two (three) times the pion mass, respectively.
(For a discussion of this issue see, for instance, \cite{rho_models2}.) 
If a $\rho$ meson is generated at normal nuclear matter density $n_0$
its mass is distributed in 
accordance with the spectral function (see Eq.~(\ref{mass_decay}) below). 
If the meson propagates into a region of higher density then the 
mass will be lowered according to 
the action of ${\rm Re} \Sigma^{ret}$ in Eq.~(\ref{eq:m}). However
if the meson comes near the threshold 
the width $\hat{\Gamma}$ becomes small and the second term of the right
hand side of Eq.~(\ref{eq:m}) dominates and reverses this trend leading to an 
increase of the mass.

The life time of unstable particles is also accessible in the framework 
of the transport equations for resonances. As it was shown in Ref.~\cite{Leupold01}
this description leads to a life time $\tau = d\delta / dE$, where $\delta$ is the
energy dependent scattering phase in the formation or the respective decay of the resonance.
Although this relation is known for a long time \cite{baz} it was introduced
only recently in the context of a BUU transport treatment \cite{Danielewicz}.
This prescription is very different from the commonly
used formula $\tau = \hbar/\Gamma$. Especially if the resonance is a $p$ wave
resonance the life time tends to small values near the threshold in the former
case, while it approaches large values in the latter one. 
If the resonance decays into several channels, the total width is the relevant
quantity which describes the phase of the amplitude common to all decay channels:
\begin{equation}
       \label{delta}
       \tan \,\delta = \frac{-\frac{1}{2}{\hat \Gamma}} 
          {p^2-m_0^2-{\rm Re} \Sigma^{ret}}.
\end{equation}
Therefore, the decay rate into a special channel $c$ is given by the partial width
$\Gamma_c = b_c \Gamma^{tot}$ according to
\begin{equation}
       \label{lifetime}
       \tau^{-1} = b_c \Gamma^{tot} = b_c (d\delta / dE)^{-1} 
       \label{decayrate}
\end{equation}
with $b_c$ being the branching ratio of the decay into channel $c$.
If we do not mention otherwise, we use the standard prescription for the life
time, but in some cases we study the effect of using Eq.~(\ref{decayrate}), as
well.

\subsection{Particle production}

In most instances, a vector meson $V$ is created by the decay 
of a baryon resonance $R$ in the BRoBUU code.  
Thus, mesons are created in two-step processes like $NN\leftrightarrow NR$ 
with subsequent decay $R\leftrightarrow VN$. 
As mentioned above the BRoBUU model includes 24 non-strange baryon resonances. 
Their parameters (mass, width and branching ratios) are determined 
by a global fit to pion-nucleon
scattering data, while resonance production cross sections are fitted to inelastic
nucleon-nucleon scattering cross sections \cite{wolf97}. Since there are very
few $np\rightarrow RN$ data we assume that 
\begin{equation}
\sigma_{np\rightarrow RN} = a \sigma_{pp\rightarrow RN} ,
\end{equation}
where $a$ is a channel independent (except one, see below) constant, and its
value $a = 1.34$ is obtained from a fit to the few existing data. We use that
prescription for all resonances (R) except for N(1535), the main source of
$\eta$ meson, where experimental data 
indicate a much higher value of $a_{N(1535)}= 5$.

Our approach is in contrast to other
ones where individual elementary hadron reaction channels are
parameterized independently from one another. Using such coupled channel 
approach could allow us to obtain cross section for not or poorly measured
channels.

The in-medium spectral functions of $\omega$ and $\rho$ mesons also have to be
taken into account when their test particles are created.
In the resonance decay the mass distribution of the generated test particles for 
mesons results from an interplay of phase-space effects and the in-medium
spectral functions ${\cal A}$ of the created meson. For the decay of
a resonance of mass $m_R$ in a meson of mass $m$ and a baryon of mass $m_N$ 
we use the phase space distribution in the final state with a 
constant matrix element squared
$\vert {\cal M} \vert^2$
\begin{equation}
\Gamma =  {\cal N}\, \int d^4p_N\,\delta(p^2_N - m^2) \int d^4p_V\, 
\frac{1}{2\pi} {\cal A}(p_V) \vert {\cal M} \vert^2
\end{equation}
from which the distribution
\begin{equation}
\label{mass_decay}
\frac{dN^{R \to N V}}{dm_V} = {\cal N}\,  \,m_V \, 
\lambda^{1/2}(m_R^2,m_N^2,m_V^2) \, {\cal A}(m_V)
\end{equation}
results, where  $\lambda$ is the triangle function 
$\lambda (a^2,b^2,c^2) = (c^2-a^2-b^2)^2-(2ab)^2$.
${\cal N}$ is an appropriate normalization factor.

We also include meson emission during a transition $R \to R' V$ from  a resonance 
state  $R$ to another resonance $R'$ 
with $R' \, =\, \Delta(1232)$, $N(1440)$, $N(1520)$, $N(1535)$.

\begin{figure}[!htb]
\hspace*{-1mm}
\centerline{
\epsfig{file=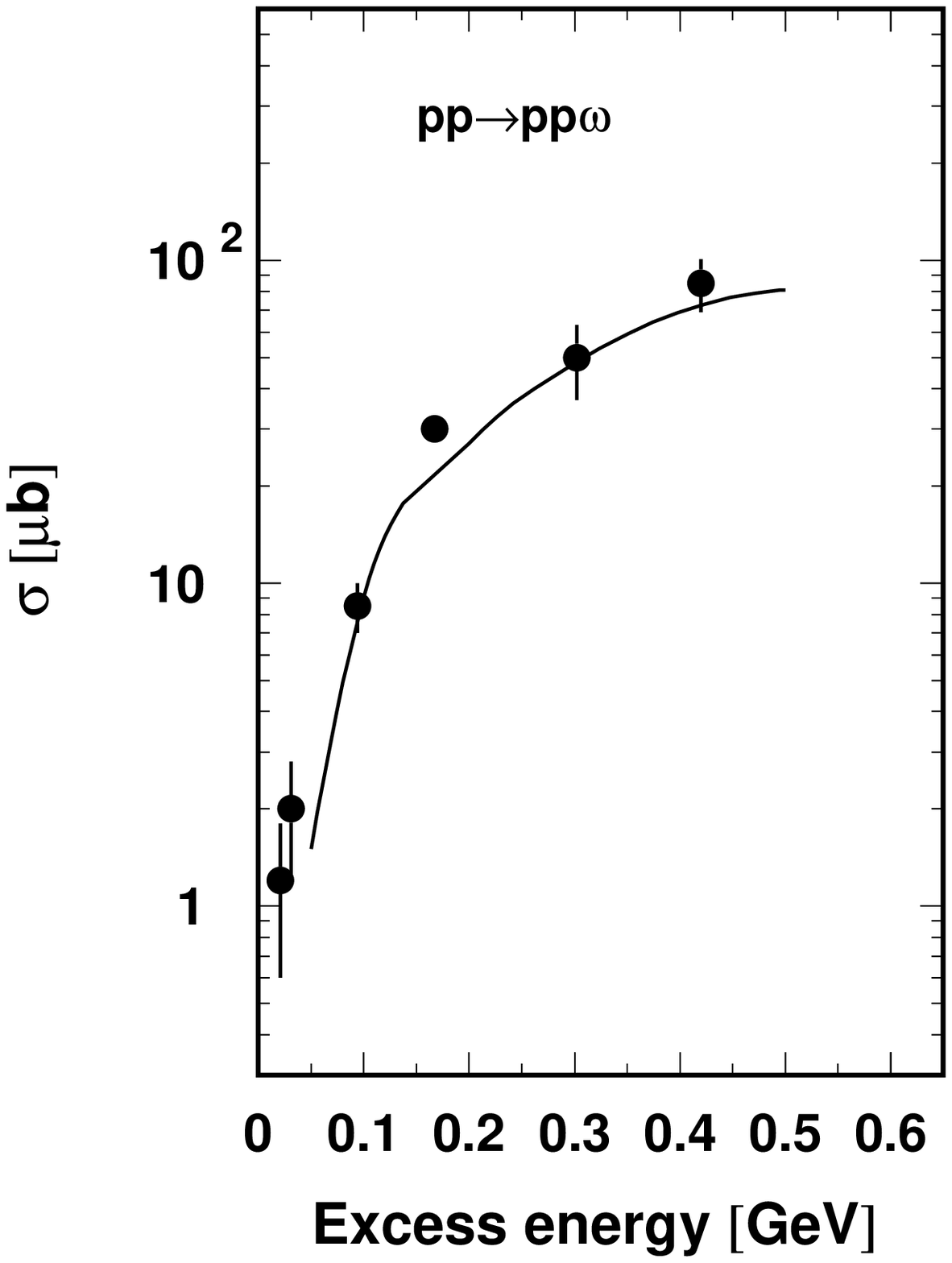,width=0.3\linewidth,angle=0} \hspace*{1mm}
\epsfig{file=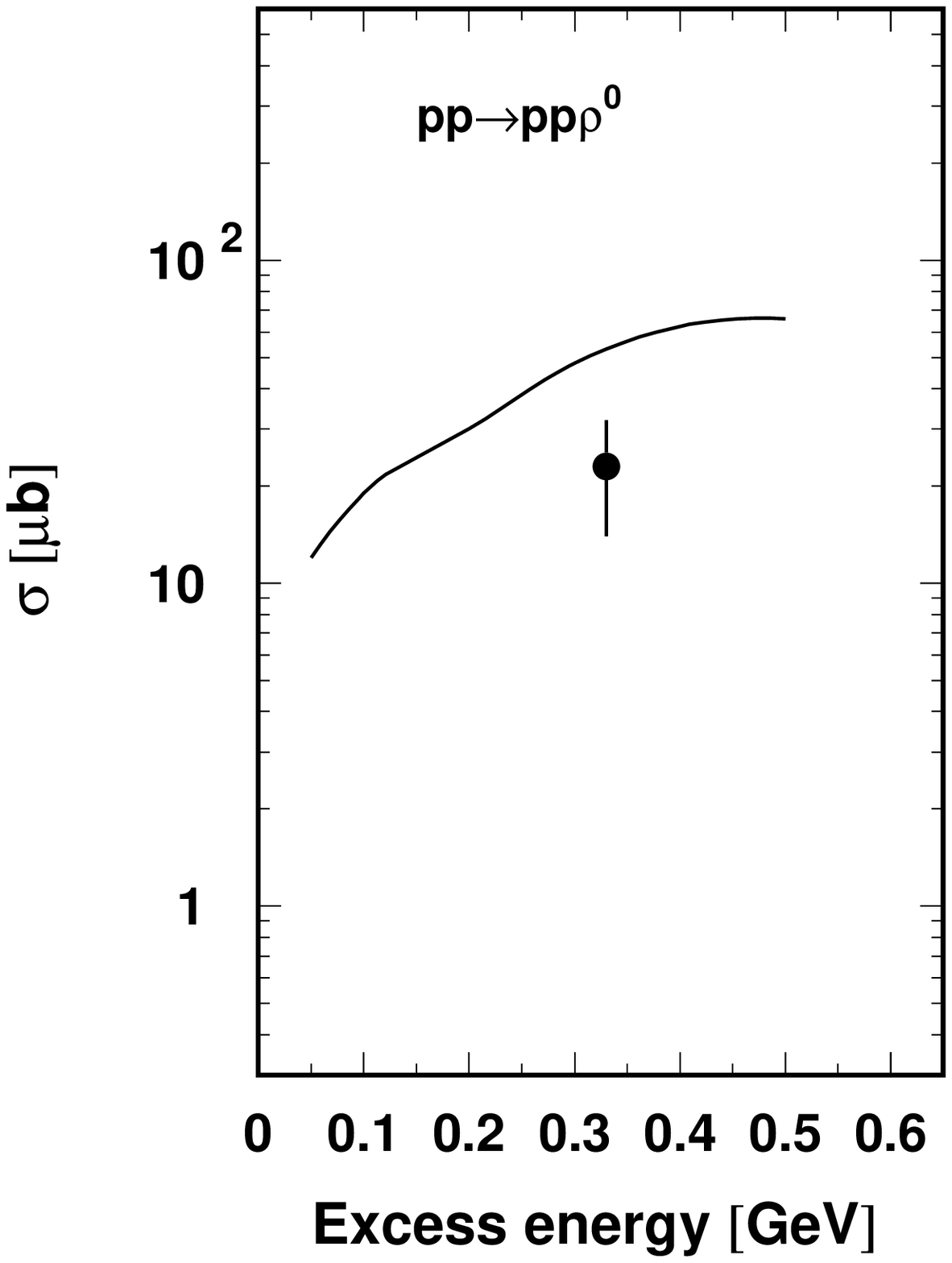,width=0.3\linewidth,angle=0} \hspace*{1mm}
\epsfig{file=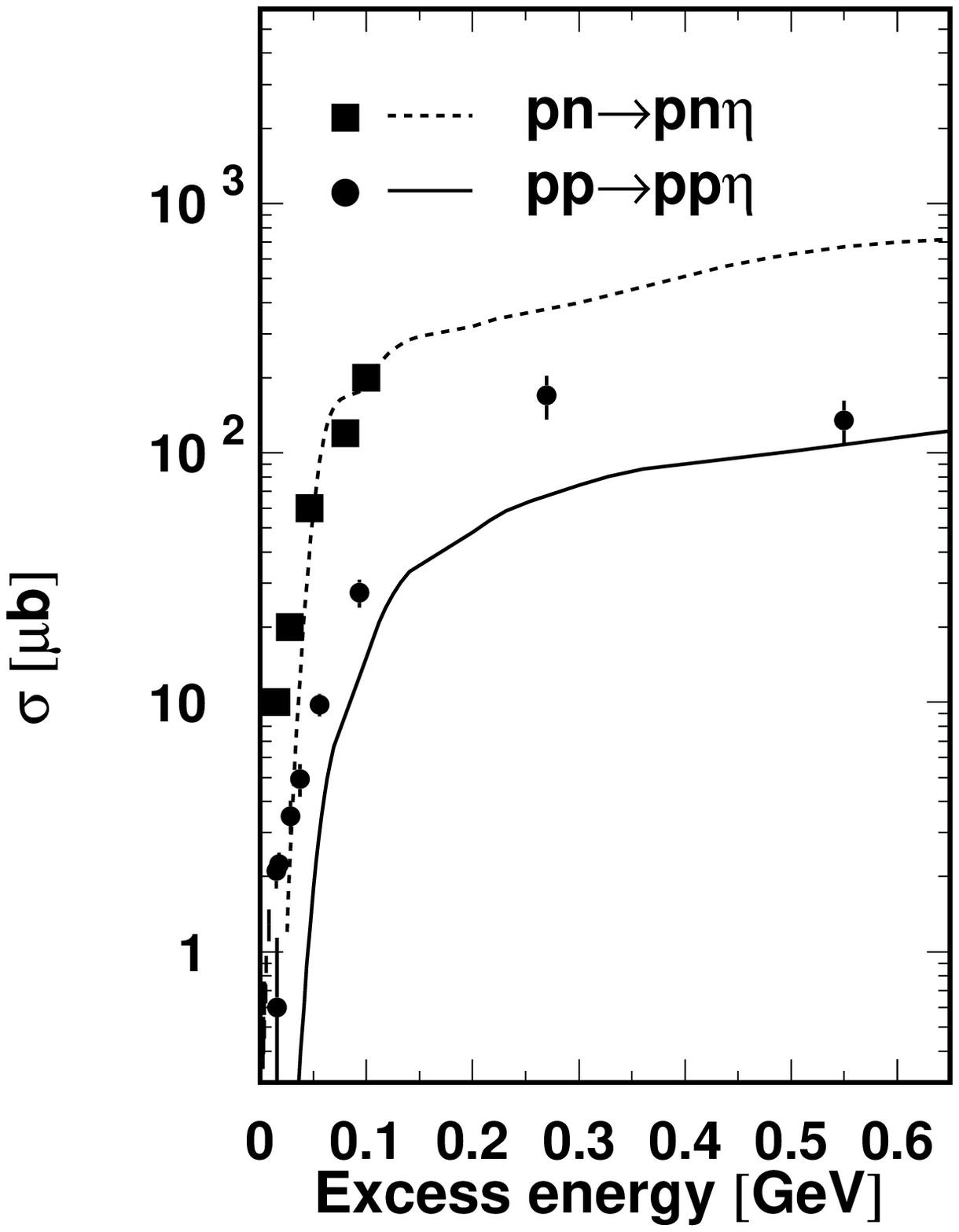,width=0.3\linewidth,angle=0}
}
\vspace*{-1mm}
\caption{\it Production cross sections of $\omega$ (left), 
$\rho^0$ (middle) and $\eta$ (right) mesons in $pp$ collisions
as a function of the excess energy in comparison with data 
\cite{SATURN,COSY,DISTO,HERA,DISTOrho,Calen,Hibou,Smyrski,Chiavassa}.}
\label{fig_ome_prod}
\end{figure}

In Fig.~\ref{fig_ome_prod} we compare the cross section calculated with
our parameter set  with data measured by the collaborations SATURN, COSY and DISTO
\cite{SATURN,COSY,DISTO,HERA,DISTOrho,Calen,Hibou,Smyrski,Chiavassa}.
The relevant range region for collisions in the 1 - 2 AGeV region is 
at excess energies below 0.5 GeV. We recognize that 
the $\omega$ production in $pp$ collisions is well reproduced by our 
model parameters. The $pn \to pn \omega$ cross sections are about 1.5 times 
larger than the $pp$ cross sections.
The one-boson exchange model in \cite{Kaptari_omega} predict even a ratio of two.

For the $\rho^0$ production near threshold there are not many $pp$ measurements 
near threshold.  Measurements are hampered by the large $\rho$ width which
make it difficult to discriminate $\rho$ mesons from sequential two pion emission.
At an excess energy of 0.33 GeV a $\rho^0$ 
cross section of  23$\pm$9  $\mu$b \cite{DISTOrho}
has been measured, where $\rho^0$ mesons were identified
by pion pairs with with masses above 0.6 GeV. The here employed 
global fit including  many elementary channels
overestimates this cross section by a factor of 2.3.
As a consequence we will reduce the $\rho$ production by this 
factor in the following.

With respect of $\eta$ production our model describes well the
production in $pn$ collision, however seems to underestimate the production in
$pp$ collisions. Since the cross sections in $pp$ collisions are anyhow smaller
than those of $pn$ collisions, this fact will not seriously affect our final results.

Furthermore the $\rho$ mesons can also be created in pion annihilation
processes $\pi + \pi\to\rho$ (see below).

\subsection{Di-electron production}

The di-electron production from direct vector meson decays $V \to e^+ e^-$
is calculated  by integrating the local decay probabilities 
along their trajectories in accordance with Eq.~(\ref{decayrate}).
The branching ratios $b_c$ of the vector mesons are taken from
experimental data at their pole masses. The mass dependence of 
this branching ratio is assumed to behave proportional to $m_V^{-3}$ in accordance
with the vector meson dominance model.

The subleading so-called direct channel
$\pi \pi \to \rho \to e^+ e^-$ is treated with the $\rho$ meson
 formation cross section
\begin{equation}
\sigma (M) = \frac{\pi}{3p^2} 2m_\rho^0 \, \Gamma(p) \, {\cal A}_\rho
\end{equation}
with $m_\rho$ being the pole mass of the $\rho$ meson and $\Gamma_\rho(p)$
the vacuum width of the $\rho$ resonance. The in-medium effects are encoded 
in the spectral function ${\cal A} $.

The $\rho$ meson produces di-electrons with a rate of
\begin{equation}
\frac{dN_{e^+e^-}}{dt} = \left( \frac{m_\rho^0}{m_\rho} \right)^3 \, b_c \, \Gamma(p). 
\end{equation}

We also include into our simulations a bremsstrahlung contribution which is guided
by a one-boson exchange model adjusted to $pp$ virtual bremsstrahlung 
and transferred
to $pn$ virtual bremsstrahlung \cite{Kaptari}. Actually, we use
\begin{equation}  
\label{brems}
\frac{d \sigma}{d M} = \frac{\sigma_\perp}{M} \, 
\frac{\alpha^2}{6 \pi^3} \int  \frac{d^3q}{q_0^3}  \frac{R_2(\bar s)}{R_2(s)}. 
\end{equation}
Here $M$ is again the $e^+e^-$ invariant mass, $R_2$ denotes 
the two-particle phase space volume, $\sqrt{s}$ stands for the c.m.s.\
energy, $\bar s$ is the reduced energy squared
after the di-electron emission, and $\sigma_\perp (s)$ is the
transverse cross section. Equation~(\ref{brems}) can be approximated:
\begin{equation} \label{sigbrems}
\frac{d \sigma}{d M}  = \frac{\alpha^2}{3\pi^2} \frac{\sigma_{tot}}{M}\,
\frac{s-(m_1+m_2)^2}{e_{cm}^2} \Big[ \ln(\frac{q_{max}+q_{0max}}{M}) -
\frac{q_{max}}{q_{0max}} \Big] .
\end{equation}
This approximation is applied to $pn$ and $\pi$N collisions using the 
respective corresponding total cross sections $\sigma_{tot}$;
$e_{cm}$ stands for the energy
of the charged particle in the rest system of the colliding particles 
with masses $m_1$ and $m_2$, $q_{0max}=(s+M^2-(m_1+m_2)^2)/2\sqrt{s}$ 
is the maximum di-electron energy, and $q_{max}=\sqrt{q_{0max}^2-M^2}$
denotes the maximum di-electron momentum.
It should be noted, however, that this cross section is still rather uncertain.

An essential di-electron contribution comes from the 
Dalitz decays of $\pi^0$, $\eta$, $\omega$ mesons and the excited 
baryon resonances emitting a di-electron together with a photon or nucleon.
The decay rate $\Gamma_{Dal}$ for a di-electron of mass $M$
for mesons can quite generally brought into the form
\begin{equation}
\frac{d\Gamma_{Dal}}{dM} = \frac{4\alpha}{3\pi M}  \Gamma_\gamma
         \left(1 - \frac{M^2}{m_V^2} \right)^3 \, F(M)^2. 
\end{equation}
The Dalitz decay rates are assumed to be given by the photon partial $\Gamma_\gamma$
width which have been taken from experiment \cite{PDG}.
The relevant form factors $F(M)$ for the mesons being considered 
are summarized in \cite{Ernst}.

\begin{figure}[!htb]
\hspace*{-1mm}
\centerline{
 \epsfig{file=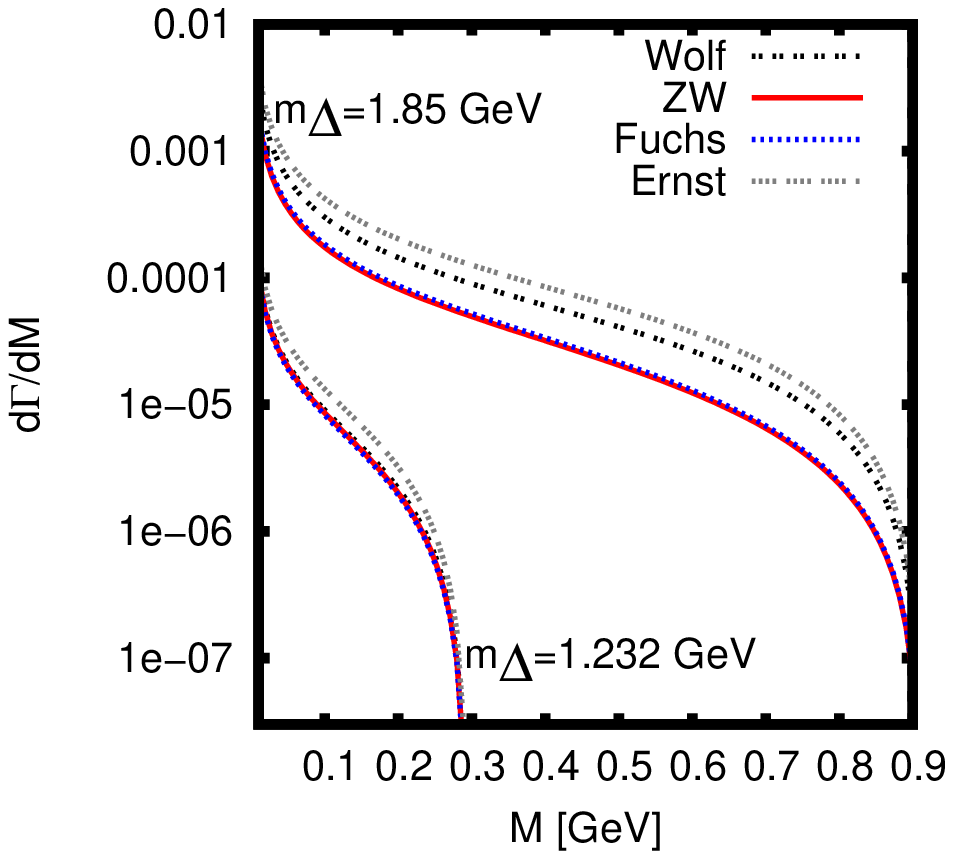,width=0.45\linewidth,angle=0} 
 \epsfig{file=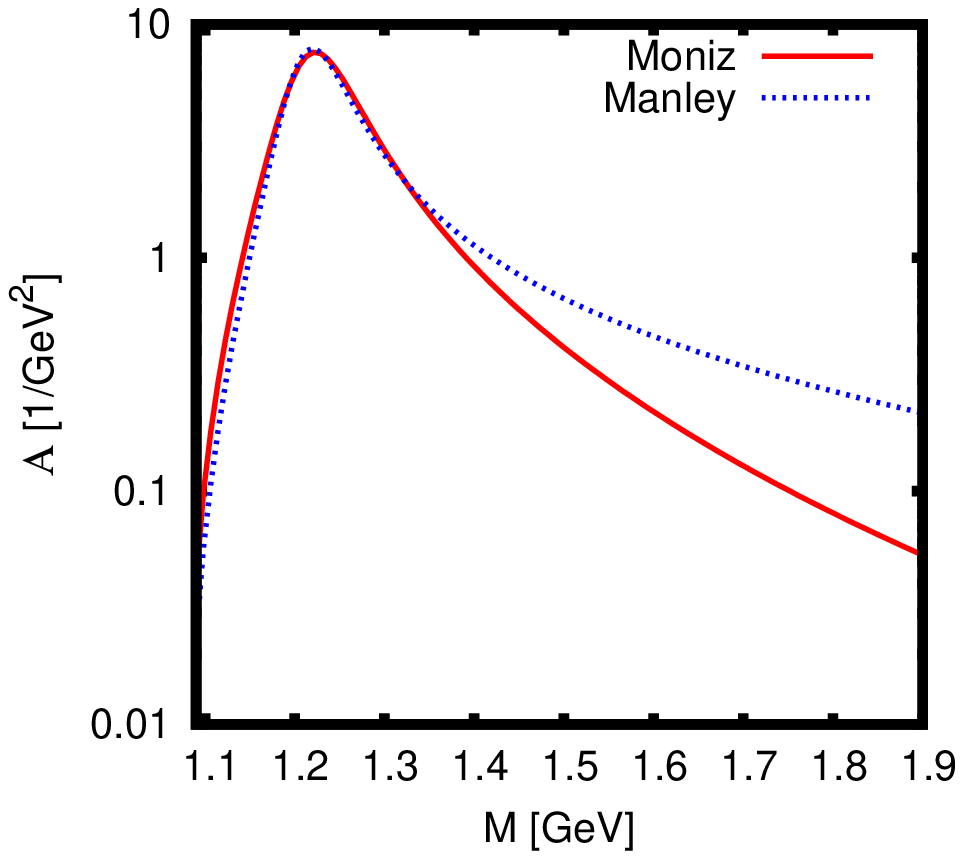,width=0.45\linewidth,angle=0}}
\vspace*{-4mm}
\caption{\it Dalitz-decay of $\Delta(1232)$ for two energies
(masses) $m_\Delta$: 
1.232 GeV and 1.85 GeV with the description from above
\cite{Wolf90,pppaper,Fuchs,Ernst} (left panel) and the spectral function with 
two different cut-off prescriptions (right panel).
See text for further details.}
\label{Delta-Dalitz}
\end{figure}

The Dalitz decay of the baryon resonances is treated as in \cite{pppaper}.
The most important contribution to the di-electron spectra of these come from 
the $\Delta(1232)$ resonance.
There are also other models \cite{Ernst,Wolf90,Fuchs} for the Dalitz-decay. 
As can be seen on the left panel of Fig. \ref{Delta-Dalitz}, 
these models agree very well for resonance decays from the peak 
mass, however they differ substantially for $\Delta$ resonances with energies (masses) 
relevant for studying the vector meson region. There is another uncertainty 
concerning these high-energy (mass) $\Delta(1232)$ resonances. The width and 
consequently the spectrum of these resonances are sensitive on the cut-off for 
high masses. Here we show two possible parameterizations: one from Moniz \cite{Moniz}
and the other one from Manley used in the Particle Data Book \cite{PDG}.

The number of $\Delta(1232)$ resonances at energy (mass) around 1.85 GeV may depend 
on the cut-off prescription by a factor of 3. Their Dalitz decay 
(see left of Fig.~\ref{Delta-Dalitz}) panel may differ by a factor of 4. 
So the Dalitz decay 
contribution of the $\Delta(1232)$ resonance is uncertain by more than an 
order of magnitude in the vector meson region. This uncertainty may only be 
clarified by a detailed comparison of the calculation for 
$pp \rightarrow pp e^+e^-$ with forthcoming experimental data. 
Using their angular dependence one can
localize the different channels and then fix their magnitude.
Here we would like to mention that different groups use different prescription
for that channel which it is one reason why the predictions,
especially for the $\Delta(1232)$ Dalitz decay contribution, are different. 

\section{Results for 2 AGeV}

\subsection{Spectral function dynamics}

\begin{figure}[!b]
\hspace*{-1mm}
\centerline{
 \epsfig{file=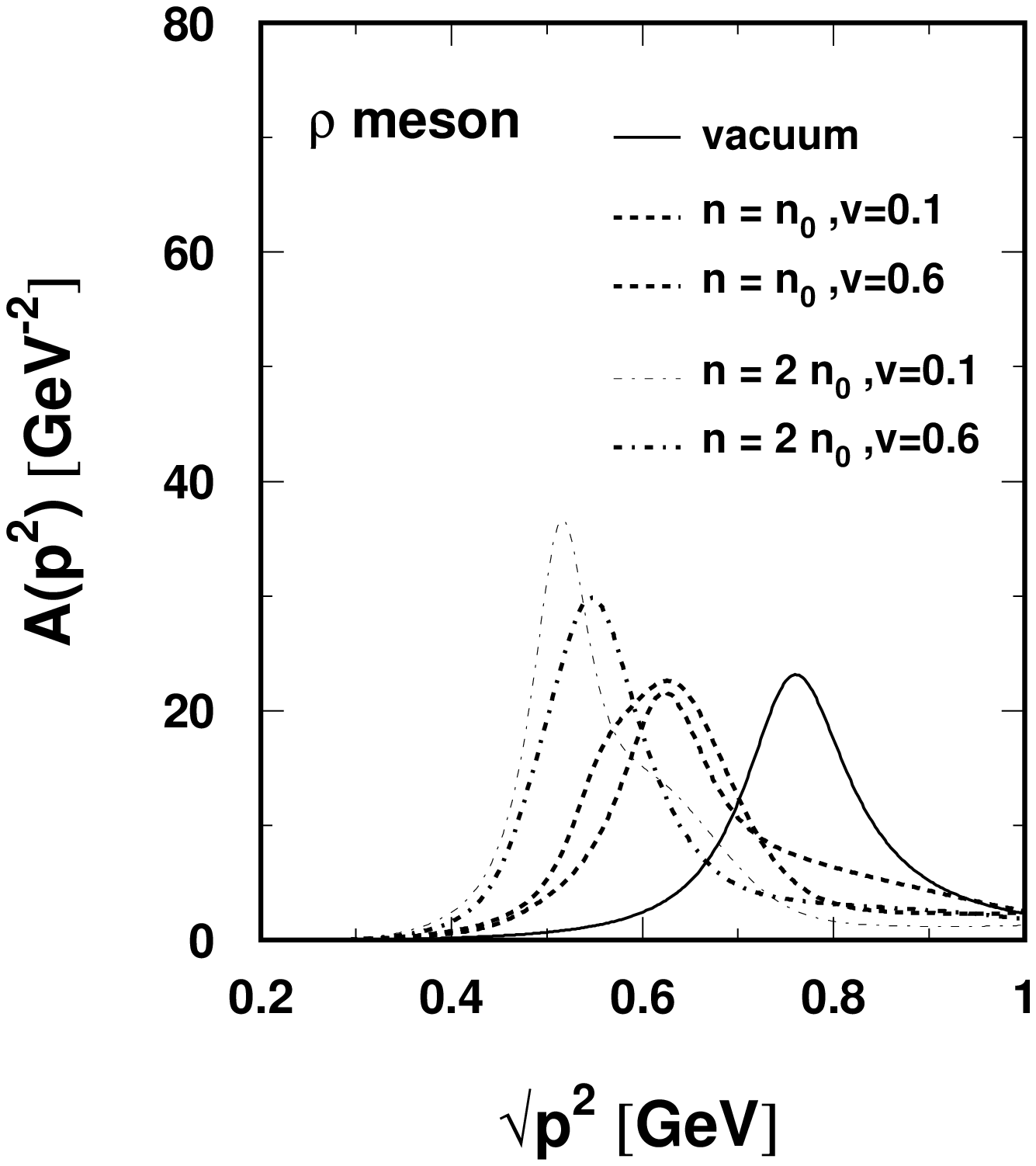,width=0.35\linewidth,angle=0} \hspace*{6mm}
 \epsfig{file=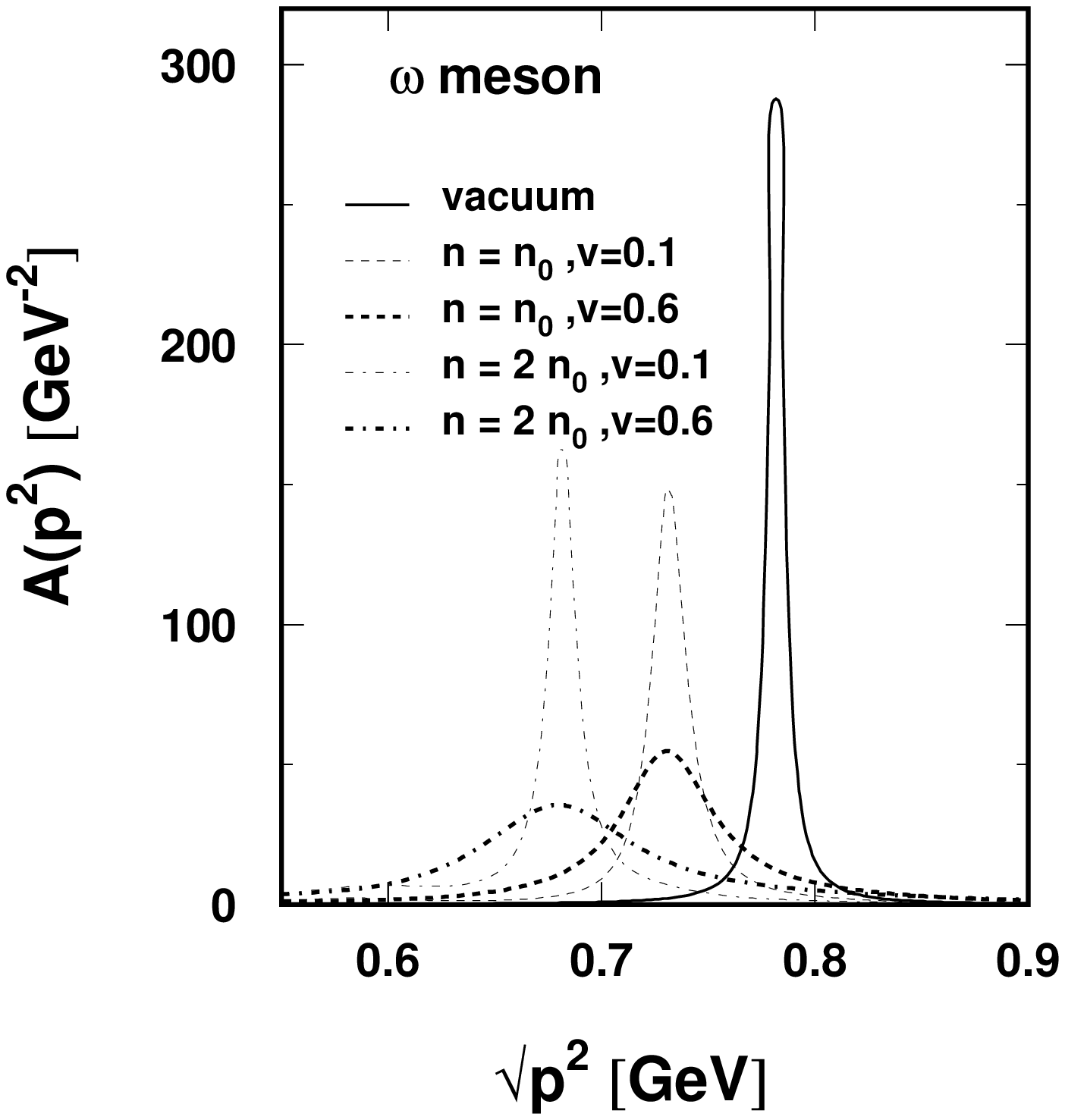,width=0.35\linewidth,angle=0}}
\vspace*{-1mm}
\caption{\it  Influence of medium effects on the spectral 
functions for $\rho$ (left panel) and
$\omega$ (right panel) mesons at various nuclear matter densities 
and velocities in matter.}
\label{fig_spec}
\end{figure}

We employ the above described code for the reaction C(2 AGeV) + C, where
data from HADES are at our disposal \cite{HADES_PRL}. 
In the present explorative study
we are going to contrast simulations with and without medium modifications
of $\rho$ and $\omega$ mesons to elucidate to which degree medium effects
may become visible in the light collision system under consideration.
In doing so we use fairly schematic medium effects 
(to be considered as an upper limit) condensed 
in a ''mass shift'' described by the above parameter $\Delta m_\omega = - 50$ 
MeV in Eq.~(\ref{areal}) for the $\omega$ meson. 
Previous CB-TAPS data \cite{Trnka} suggested indeed such a $m_\omega$ mass shift.
(See, however, \cite{Oset} for a critical discussion of this data.)
This problem is also investigated experimentally \cite{KEK} and 
theoretically \cite{Lutz,Ellis,Leupold_omega}
in calculating the $\omega$ spectral function.
The use of QCD sum rules \cite{QCDSR} then can be utilized to
translate this shift into a significantly larger shift for the $\rho$ 
meson (dictated essentially by the Landau damping term); we use here 
$\Delta m_\rho = - 100$ MeV. We are aware of experiments as reported in
Ref.~\cite{CLAS} 
which do not observe a noticeable shift of the $\rho$ meson excitation 
strength. Nevertheless,
several theoretical attempts are made to predict a possible $\rho$
''mass shift'' during the last decade.
Many of them predict a fairly large shift of strength of $\rho$ excitation
to lower energy \cite{Hatsuda_Lee,Brown_Rho}, see
also \cite{rho_models,rho_models1,rho_models2,rho_models4,
rho_models5,rho_models6,rho_models7}.
Thus, we keep this (presumably too large a) value to 
illustrate whether it would have a significant imprint on the observed spectra.

The spectral function for $\rho$ and $\omega$ mesons are shown
in Fig.~\ref{fig_spec} at two different 
densities of nuclear matter and two meson velocities 
in comparison with the vacuum spectral function. 
We would like to emphasize the strong velocity dependence of the widths, 
in particular for the $\omega$ mesons.
Despite the excitations of various baryon resonances the spectral functions
appear as relatively smooth distributions.

\begin{figure}[!htb]
\center
 \hspace*{-1mm}
\epsfig{file=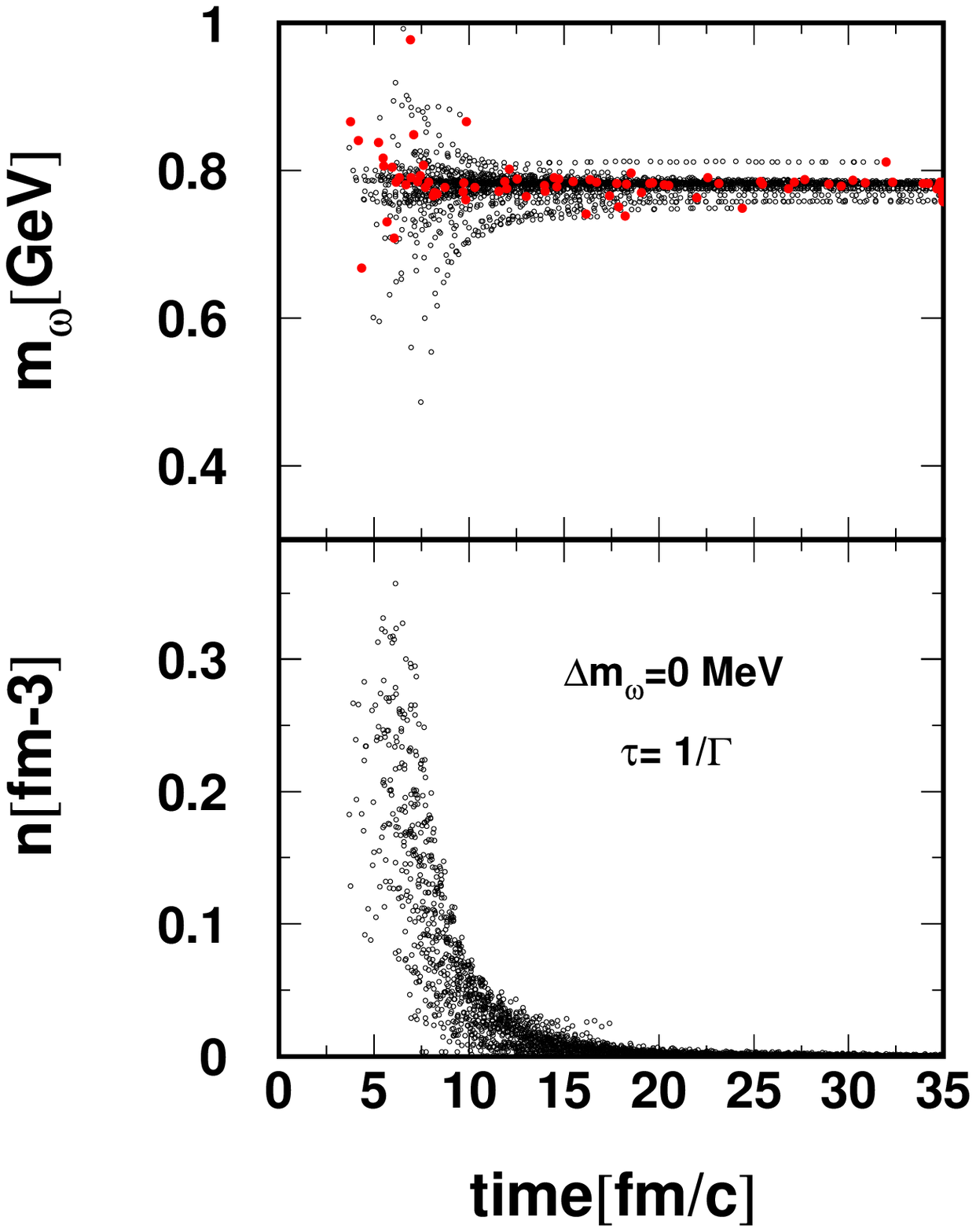,width=0.4\linewidth,angle=0} \hspace*{3mm}
\epsfig{file=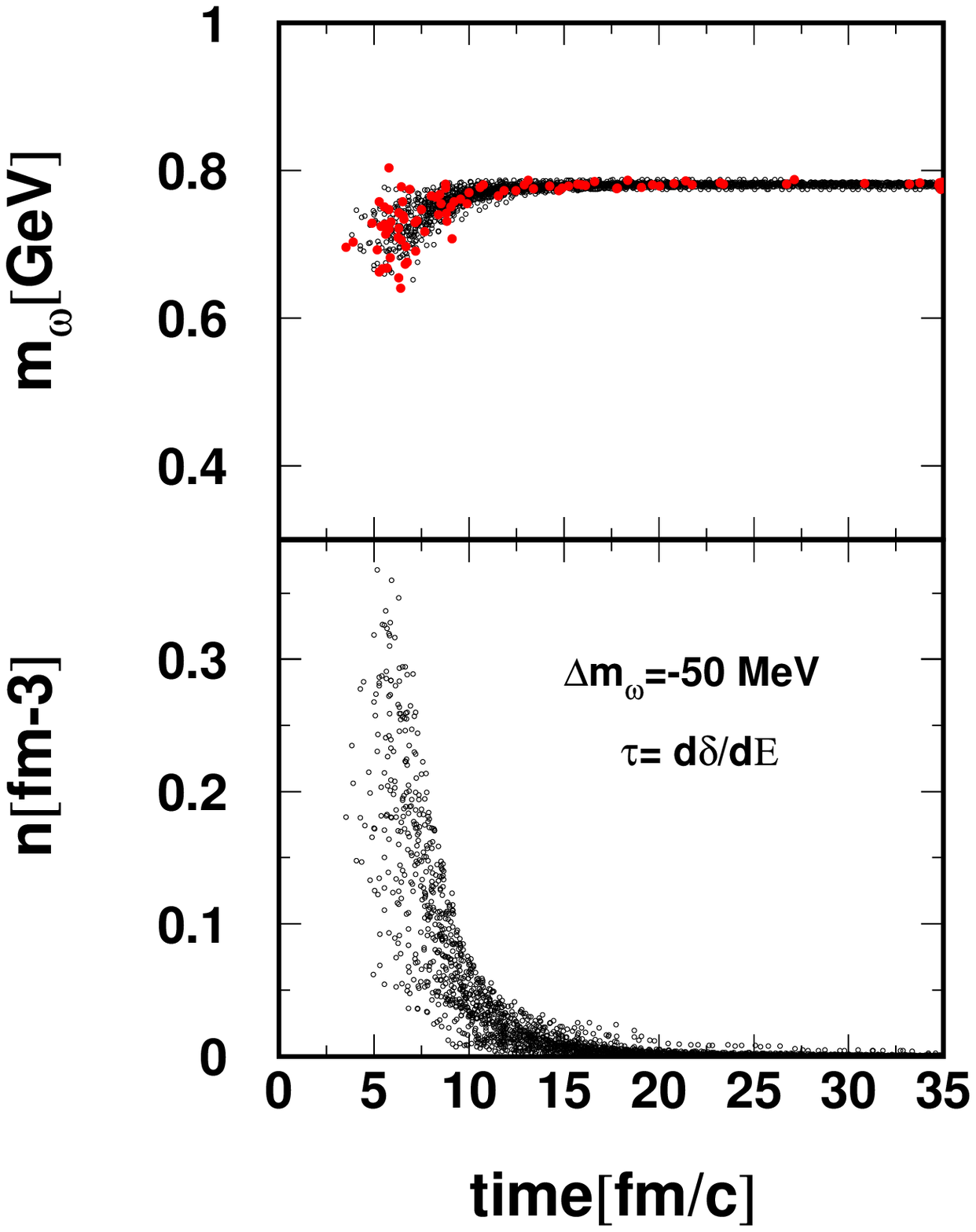,width=0.4\linewidth,angle=0}
\vspace*{-1mm}
\caption{\it Time evolution of the masses (upper panels) of about 100 test particles of
$\omega$ mesons in a C + C collision at 2 AGeV
kinetic beam energy at an impact parameter of 1 fm.
The red open circles indicate
the time instant when test particles are annihilated due to 
either pionic decay or resonance absorption.
The lower part displays the
corresponding local densities for the same test particles.
The left (without mass shift) and right (with mass shift) panels 
indicate the effect of the spectral function
and decay properties.}
\label{fig_test_ome}
\end{figure}

Let us now consider the effect of the mass evolution of the $\omega$ and 
$\rho$ vector mesons given by the 
equations of motion Eqs.~(\ref{eq:x}-\ref{eq:e}). The ensemble 
of test particles of mesons is generated in dense
matter where their masses are distributed in accordance with their broadened and
mass shifted spectral function.
In Fig.~\ref{fig_test_ome}
we show the time evolution of a small ensemble of $\omega$ test particles. 
In the calculations the nuclei touch each other at a time of 2.5 fm/c while
the density peaks at about 6 fm/c and drops at 8 fm/c below
saturation density (see lower part of the figure). 
At maximum density most of the vector mesons are created,
afterwards the mass distribution gets narrower. 
Only a few of the $\omega$ mesons decay in the dense phase 
where their masses deviate strongly
from the pole mass value. If a low density is reached the vacuum
spectral function dominates the di-electron decays which leads to  a sharp
peak at the pole mass. The right hand part of  Fig.~\ref{fig_test_ome}
shows the test particles with a spectral function with both the mentioned mass shift
and collision broadening, while the distribution on the left hand part 
is calculated with collision broadening only.

\begin{figure}[!htb]
\center
\hspace*{-1mm}
\epsfig{file=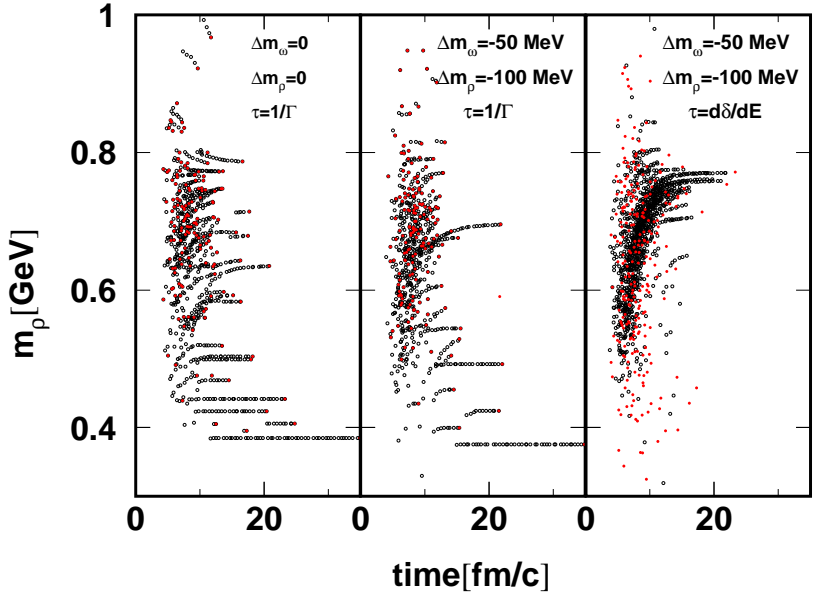,width=0.66\linewidth,angle=0} 
\vspace*{-1mm}
\caption{\it Time evolution of the masses of about 400 test particles of
$\rho$ mesons in a central C + C collision at 2 AGeV
kinetic beam energy at an impact parameter of 1 fm.
The particles in the left panel are calculated with a spectral function without
a mass shift while the middle and the right ones the indicated mass shift were used.
Note the different prescription for the life time indicated in the legends.}
\label{fig_test_rho}
\end{figure}

Figure~\ref{fig_test_rho} displays the analog behavior of the test particles of 
$\rho$ mesons. In the high density stage
one recognizes the large spread of the $\rho$ masses  
of about 300 MeV due to the imaginary part of the self-energy. 
Most of the $\rho$ mesons decay rapidly (life time $\tau < 2$ fm/c). 
If we use  the relation  $\tau=1/\Gamma$ for the life time 
then essentially the high-mass particles with their 
large width decay rapidly and, hence, have little chances to
radiate di-electrons during their short life time.  
They look like flashes occurring only in a narrow time interval,
thus not causing a longer paths of adjacent points in the mass vs.\ time plot.  
New mesons are readily created at later times and lower densities. A few 
low-mass $\rho$ mesons survive these periods and can still be found at 25 fm/c.
Therefore, one expects a shift of the di-electron spectra to lower masses.
Quite a different picture shows the right hand panel where the life time 
is calculated accordingly to Eq.~(\ref{lifetime}). 
Here the low mass particles have a shorter life time than 
the more massive ones. 

\begin{figure}[!htb]
\center
\hspace*{-1mm}
\epsfig{file=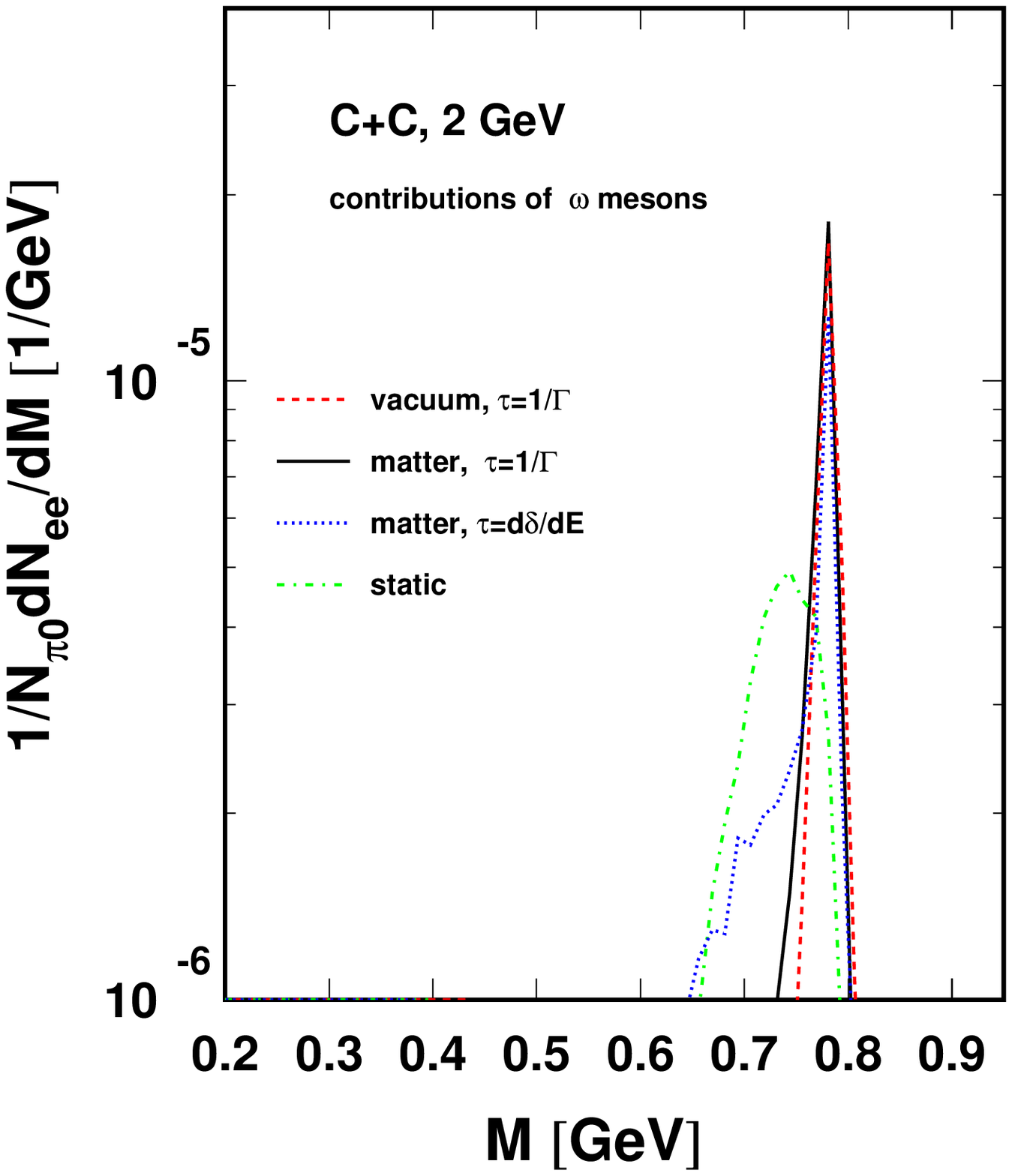,width=0.35\linewidth,angle=0}  \hspace*{9mm}
\epsfig{file=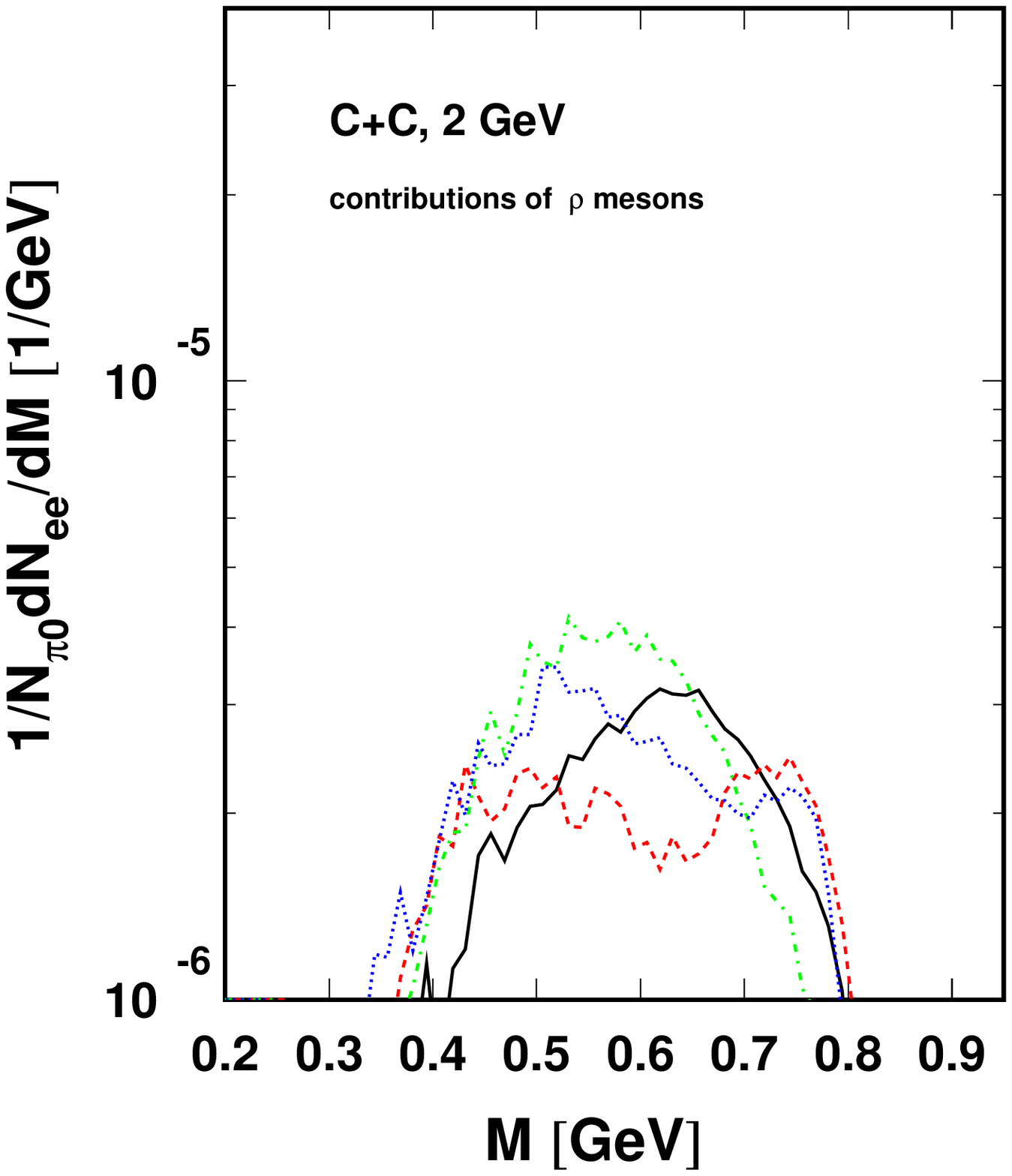,width=0.35\linewidth,angle=0}
\vspace*{-1mm}
\caption{\it Di-electron spectra 
from direct decay of $\omega$ (left panel) and $\rho$ (right panel)
mesons calculated with different assumptions for the dynamics and the spectral functions.
''vacuum'':  vacuum spectral function, 
''matter'': collision broadening and mass shifts
with two differen assumption of the life time, 
''static'': in-medium spectral function while
the mass evolution of the mesons is switched off.}
\label{fig_comp_evol}
\end{figure}

In Fig.~\ref{fig_comp_evol} we exhibit the resulting di-electron spectra
from $\rho$ and $\omega$ decays for the different prescriptions
of the life time. 
The dashed line shows the $\rho$ and $\omega$ spectra 
if the masses of the test particles are kept fixed to the
values at the instant of creation (green dash-dotted line). 
In this case the di-electron spectrum reflects the initial mass distribution
which contains the high density spectral function. 
The effect ist best visible for $\omega$ mesons near the pole mass: 
here, the peak at the pole mass
would nearly disappear when disregarding the mass evolution.
The time evolution
of the off-shell propagation pushes the resulting di-electron spectra towards their
vacuum spectral function. If the life time of the vector mesons follows
the standard expression $\tau= \hbar /\Gamma$ then the low-mass vector mesons
have sufficient time to reach their pole mass. 
This behavior is 
also clearly seen (solid line) for the $\omega$ meson. However, if
Eq.~(\ref{decayrate}) controls the decay this shift is hindered by the earlier
decay of the low mass mesons (compare solid and dotted lines).

\begin{figure}[!htb]
\vspace*{-10mm}
\center
\epsfig{file=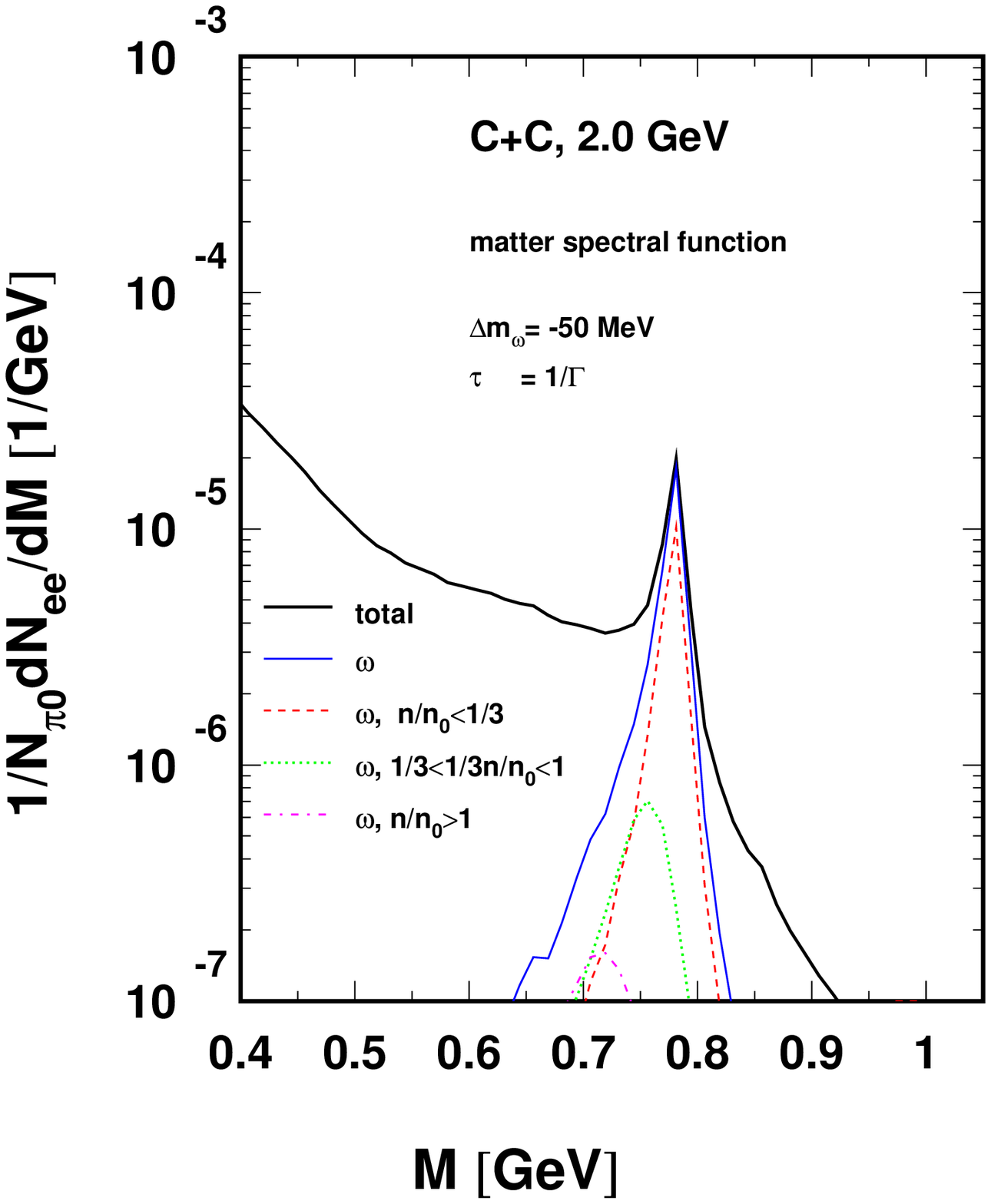,width=0.45\linewidth,angle=0} \hspace*{6mm}
\epsfig{file=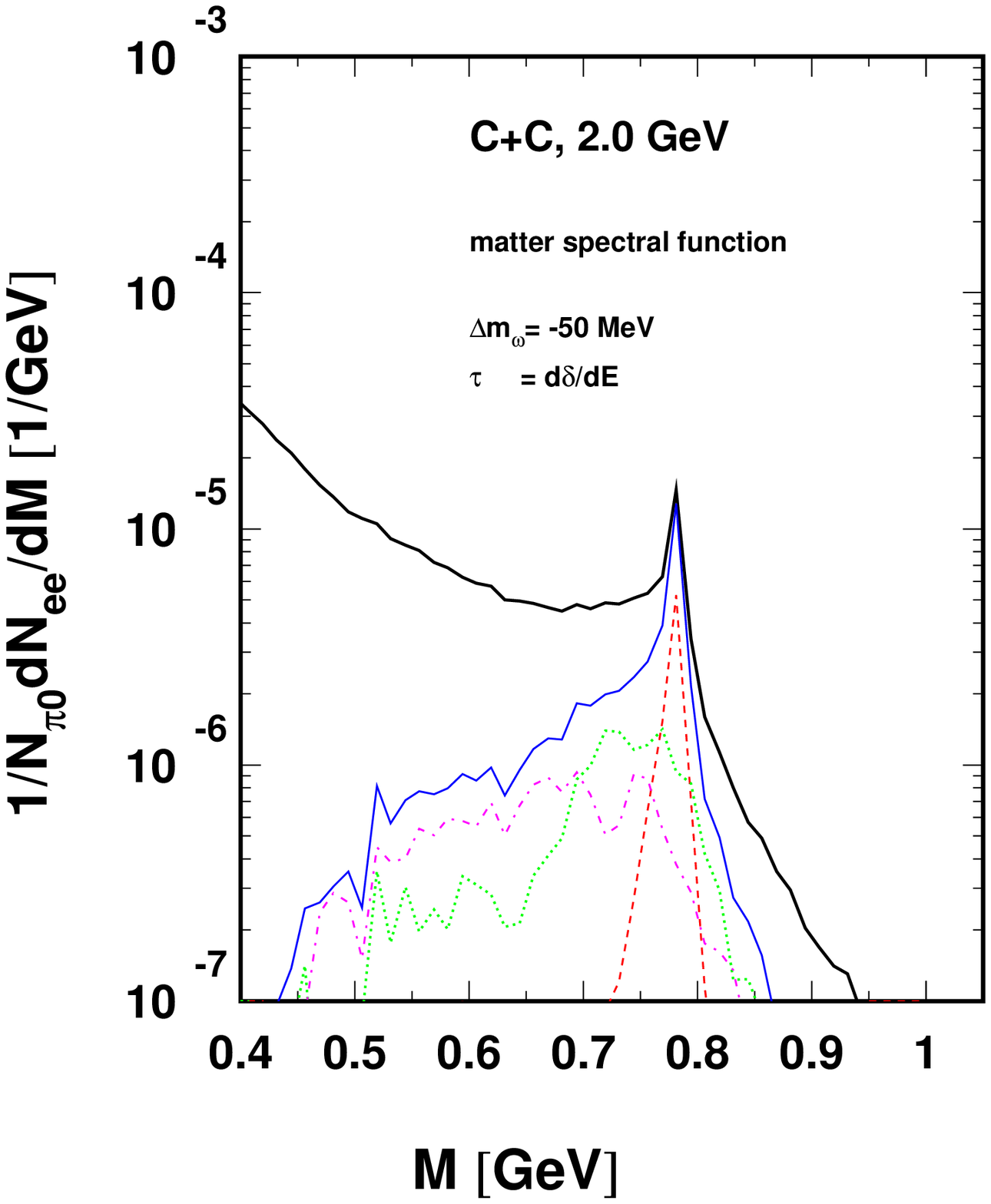,width=0.45\linewidth,angle=0}
\vspace*{-10mm}
\caption{\it
Contribution to the di-electron yield from  $\omega$ mesons in various density regions
compared with the total yield (thick solid line labelled by
''total'').  The left  picture is calculated with
the standard life time for $\omega$ mesons, while the right panel shows the effect
when  using the life time $\tau = d\delta /dE$.}
\label{fig_CC_dens}
\end{figure}

Finally we investigate the distribution of the emitted di-electrons 
as function of the density of the emitting region. We consider the effect
of the $\omega$ mesons in three different density regions: 
(i) the density $n < n_0/3$, (ii) $n_0/3<n<n_0$, and (iii) $n>n_0$. 
For the light $CC$ system di-electrons from all region have similar 
masses and can therefore hardly disentangled in experiment,
see Fig.~\ref{fig_CC_dens}.
 
\subsection{Comparison with HADES data}

\begin{figure}[!b]
\center
\epsfig{file=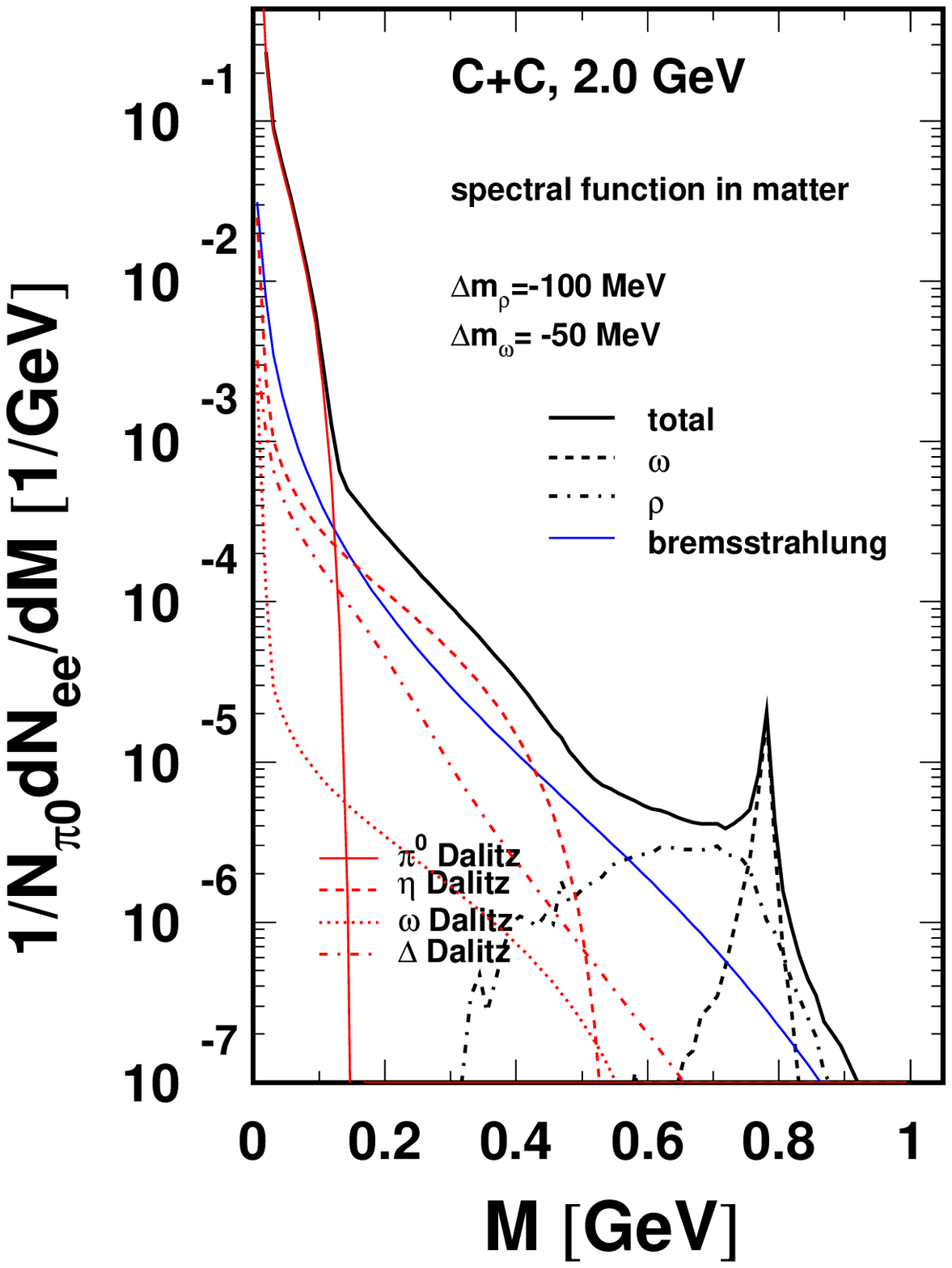,width=0.35\linewidth,angle=0} \hspace*{3mm}
\epsfig{file=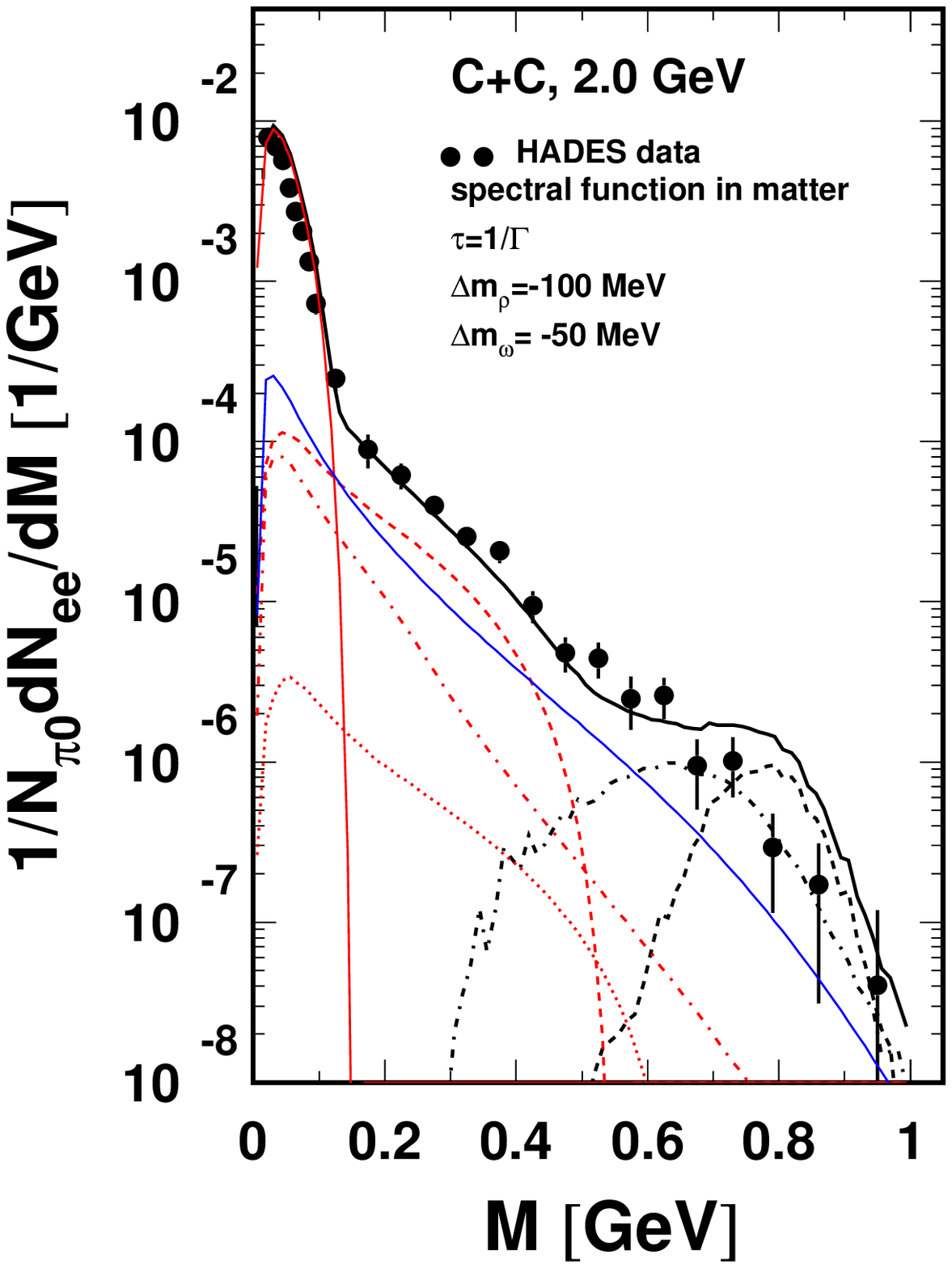,width=0.35\linewidth,angle=0}
\vspace*{-1mm}
\caption{\it Di-electron invariant mass spectra for C(2 AGeV) + C
calculated with in-medium spectral function.
Individual contributions are depicted.
Left panel: Full phase space. 
Right panel: With experimental filter
\cite{Holzmann}
and compared to HADES data \cite{HADES_PRL}.}
\label{fig_HADES_med}
\end{figure}

\begin{figure}[!htb]
\center
\hspace*{-1mm}
\epsfig{file=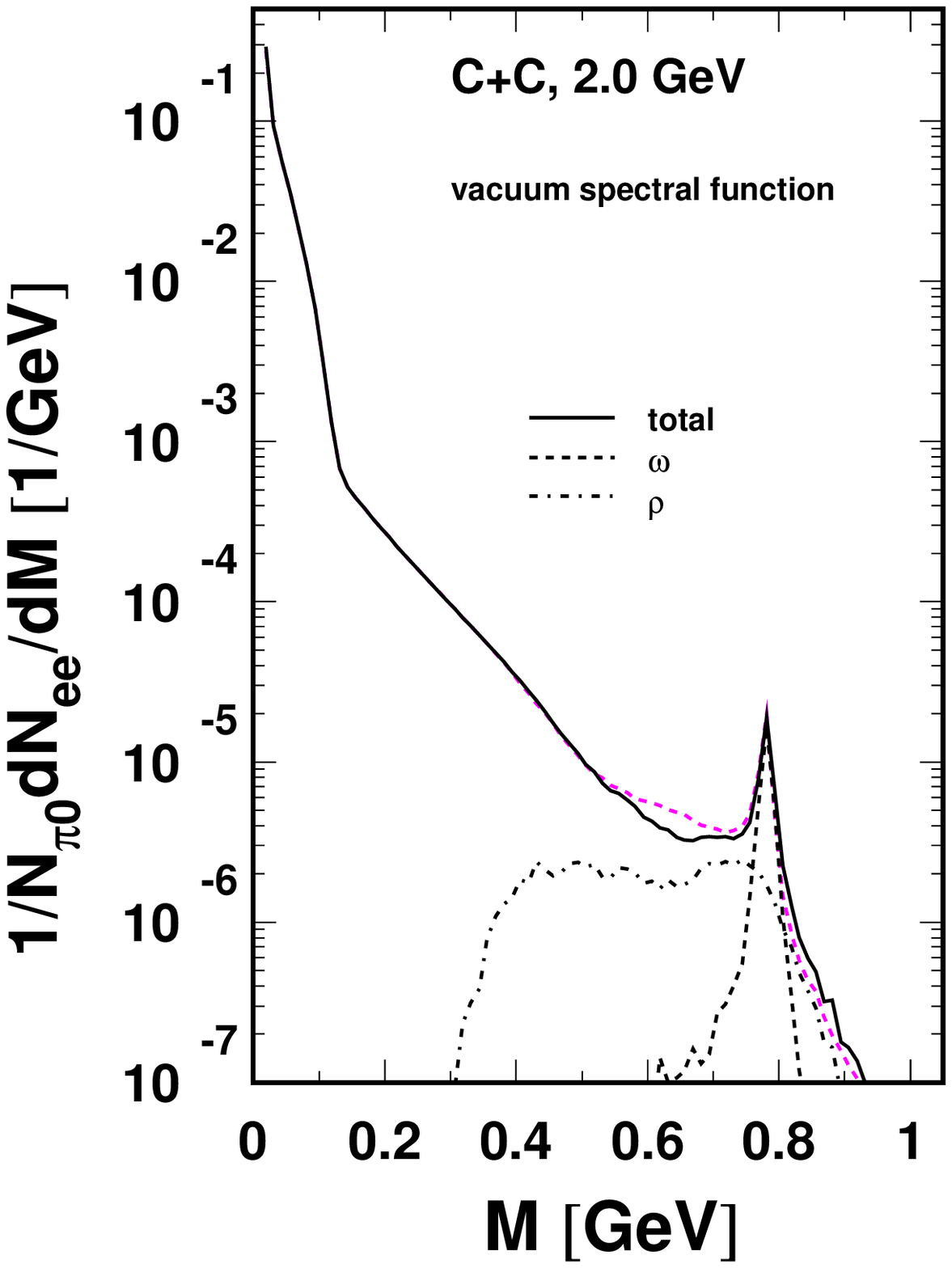,width=0.35\linewidth,angle=0} \hspace*{3mm}
\epsfig{file=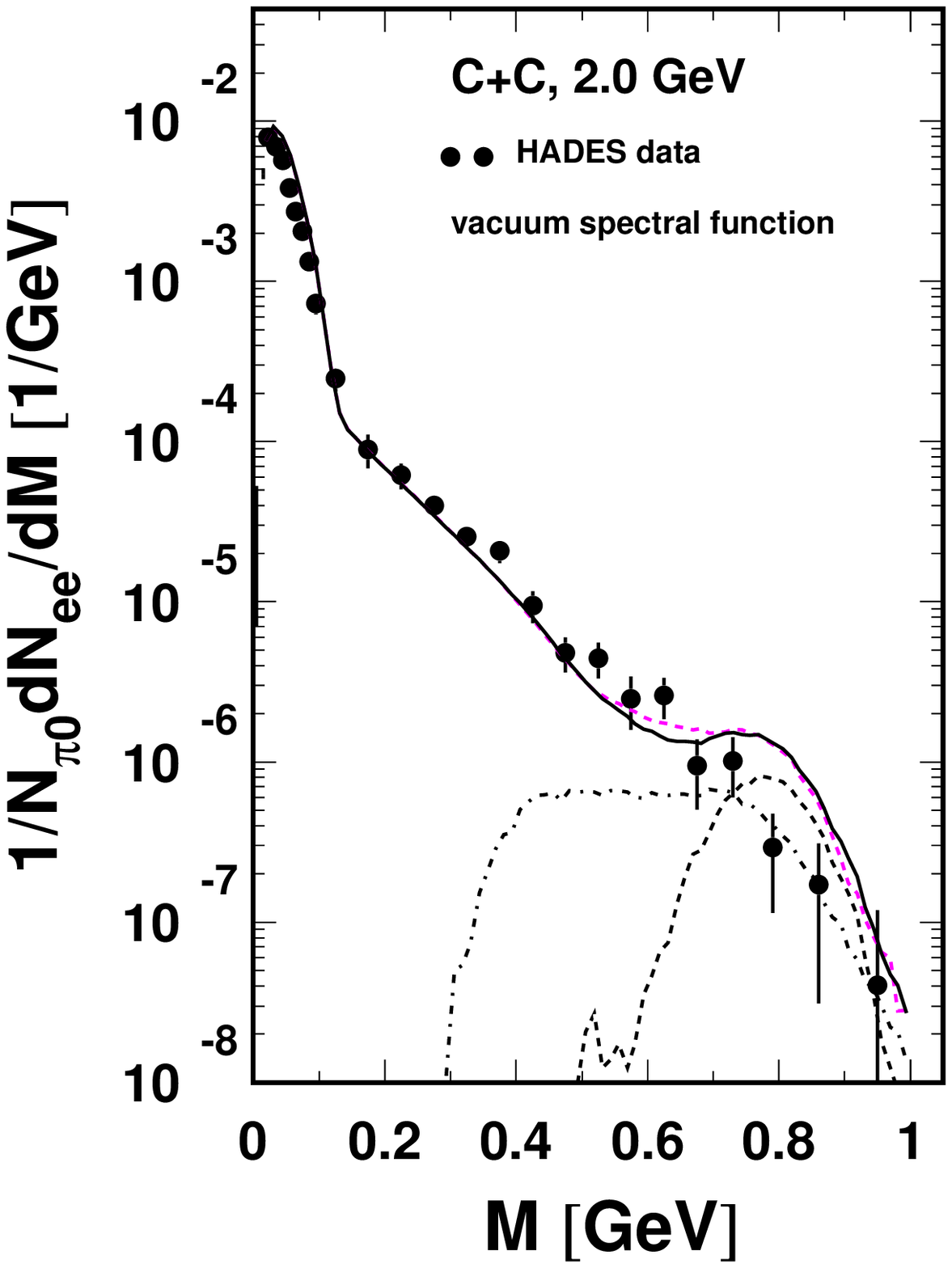,width=0.35\linewidth,angle=0}
\vspace*{-1mm}
\caption{\it As Fig.~\ref{fig_HADES_med} but with vacuum spectral function
(collision broadening is included).
The contribution from $\omega$ and $\rho$ mesons are indicated.
The violett dashed line shows the spectra calculated with the in-medium
spectral function from Fig.~\ref{fig_HADES_med}. }
\label{fig_HADES_vac}
\end{figure}

While Figs.~\ref{fig_test_ome}, \ref{fig_test_rho} and \ref{fig_comp_evol}
refer to the central point of our
work, we now look how the evolving spectral functions
compare with data. In doing so the other di-electron sources
have to be included.
The obtained di-electron spectra are represented 
in Figs.~\ref{fig_HADES_med} and \ref{fig_HADES_vac}.
In Fig.~\ref{fig_HADES_med} the results are exhibited obtained by 
including the above described pole-mass shifts as an additional medium modification 
of $\rho$ and $\omega$ mesons.
In Fig.~\ref{fig_HADES_vac} only the collision broadening is employed 
which modifies the imaginary part of the spectral function.
For comparison with the data, the HADES filter has been applied \cite{Holzmann}
accounting for the geometrical acceptance,
momentum cuts and pair kinematics. The filter causes a reduction of the strength
and a smearing of the invariant masses of the di-electrons.
The result of this filtering is always shown on the right hand panel of the figures.

In these figures we show various contributions to the 
di-electron rate. Important low-mass
di-electron sources are $\pi^0$ and $\eta$ Dalitz decays which are 
proportional to the multiplicities of their parents.
The TAPS collaboration has measured \cite{TAPS} the $\pi^0$ and $\eta$ production 
cross sections of 707$\pm$72 mb and 25$\pm$4 mb which have to be compared
to our calculations of 870 mb and 23 mb in the same reaction at the same
energy. While the values for pion production 
are overestimated the $\eta$ production is quite in agreement with the data.
(Note that the presently employed cross sections rely on a global fit of many
elementary reactions which is not optimized for special channel.) 
The Dalitz decays of $\rho$ and $\omega$ mesons and nucleon resonances 
do not contribute noticeably.

Comparing Figs.~\ref{fig_HADES_med}
and \ref{fig_HADES_vac}, the mass shifts of the vector mesons 
do not have a noticeable effect on the overall shape of the di-electron spectra  
although the peak position of the $\rho$ mesons is clearly 
shifted from 0.7 to 0.5 GeV. However, the large contribution of the cocktail
of the other sources cover the effect of the $\rho$ mesons.
Furthermore most of the $\omega$ mesons 
decay outside the dense zone and are therefore not very sensitive for 
medium effects.
Since the fine structure ($\omega$ peak) is not yet resolved in the data 
a conclusive decision cannot be made.
 
Progress could be made if the di-electron mass resolution is improved to identify 
the $\omega$ peak. However, our calculations do not point to the possibility 
of a two-peak structure (resulting from a superposition of 
vacuum decays and in-medium decays) or a
substantial smearing of the $\omega$ peak due to a density dependent shift
analog to the consideration of the $\phi$ meson (see \cite{Zschocke}).

\begin{figure}[!b]
\center
\vspace{-6mm}
\epsfig{file=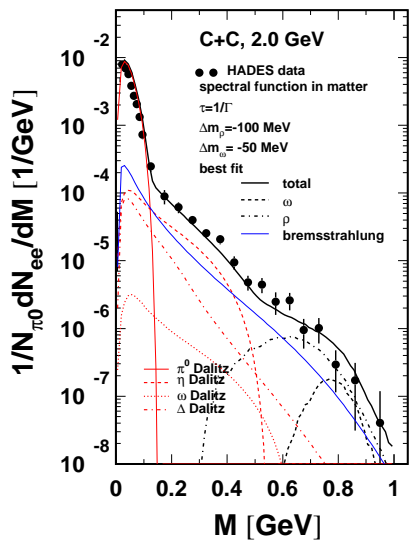,width=0.35\linewidth,angle=0}
\vspace*{-2mm}
\caption{\it
Comparison of the HADES di-electron spectrum with a calculation where the 
$\omega$ ($\rho$) cross section is scaled down by a factor 0.2 (0.8).}
\label{fig_CC_scal}
\end{figure}

The present set-up provides a reasonable description of the HADES data
\cite{HADES} for di-electron masses below 0.6 GeV. In the higher mass region
some overestimation of the data is recognized. With respect to the uncertainties
of the cross section $pp\rightarrow pp\rho$ at threshold and generally the $pn$
channel as well as the role of the resonance channels one we could try to 
improve the agreement by rescaling the $\rho$ and $\omega$ contributions.
In doing so we assume  that the spectral shapes are unaltered. To get a 
better agreement with data one needs to decrease artificially the cross sections for 
vector meson production by factors 0.2 for $\omega$ and 0.8 for 
$\rho$. Figure \ref{fig_CC_scal} exhibits this ''optimized''
comparison with data.

\begin{figure}[!htb]
\center
\vspace*{-22mm}
\hspace*{-11mm}
\epsfig{file=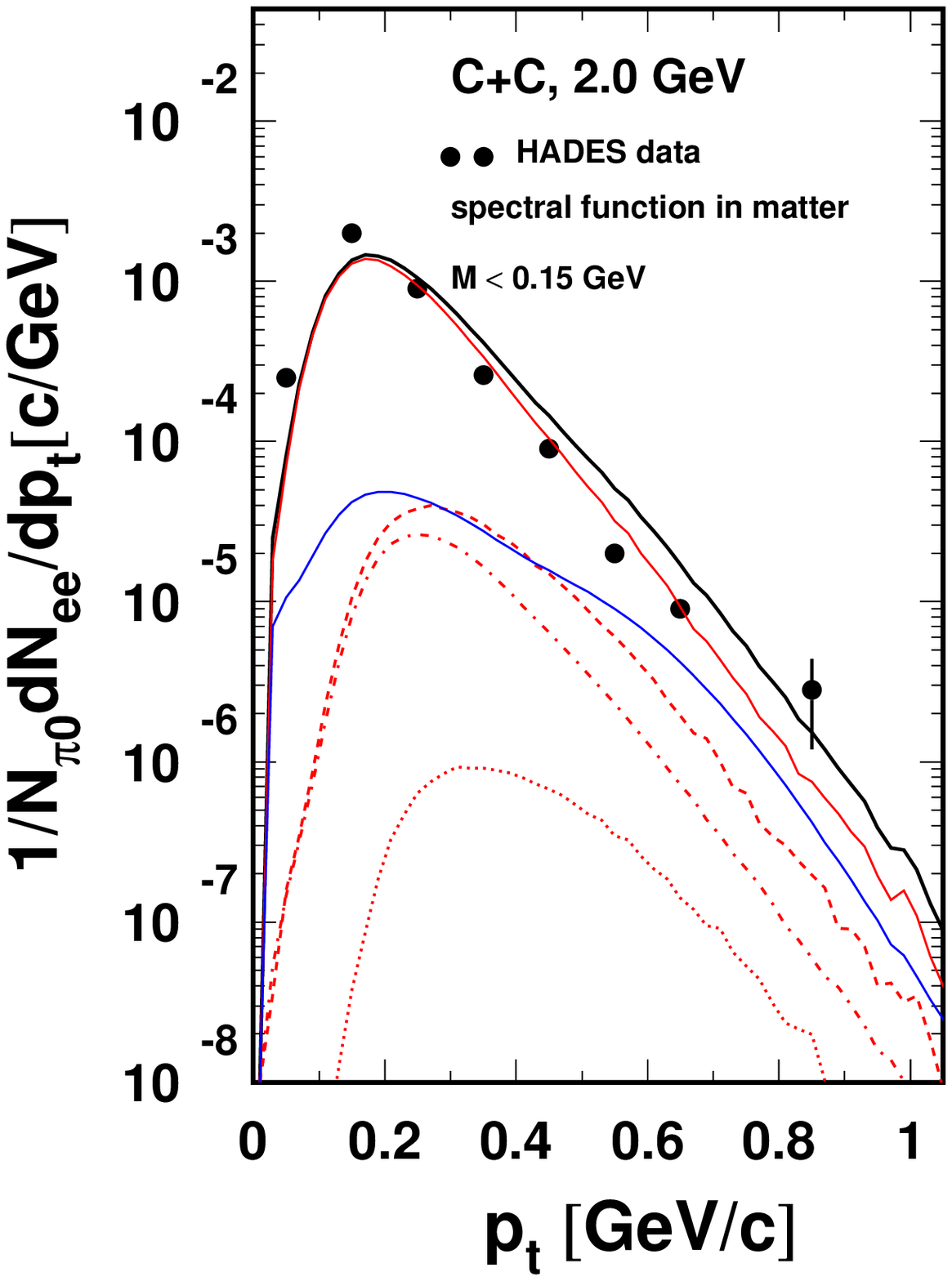,width=0.35\linewidth,angle=0} \hspace*{-17.3mm}
\epsfig{file=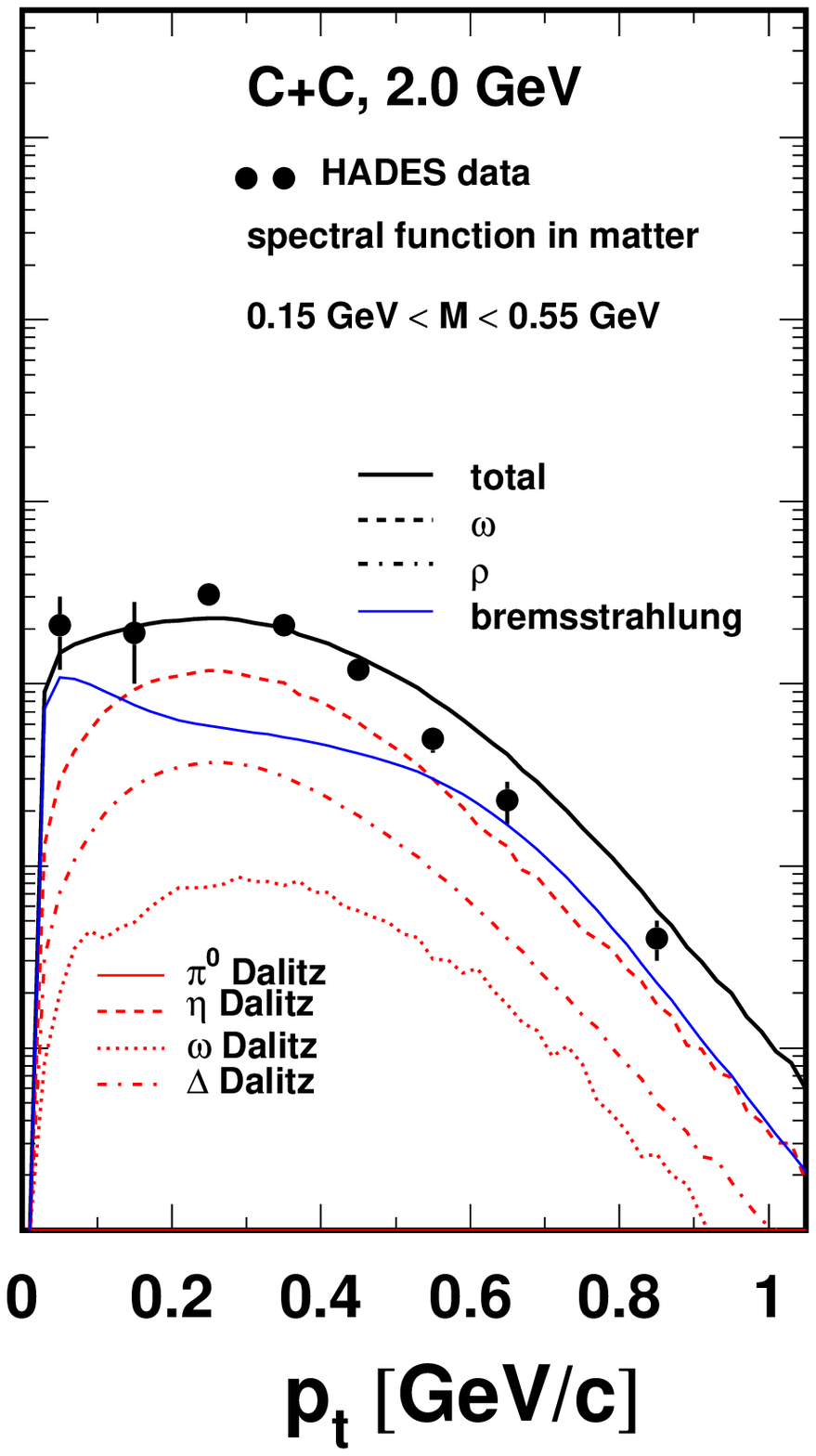,width=0.35\linewidth,angle=0} \hspace*{-17.3mm}
\epsfig{file=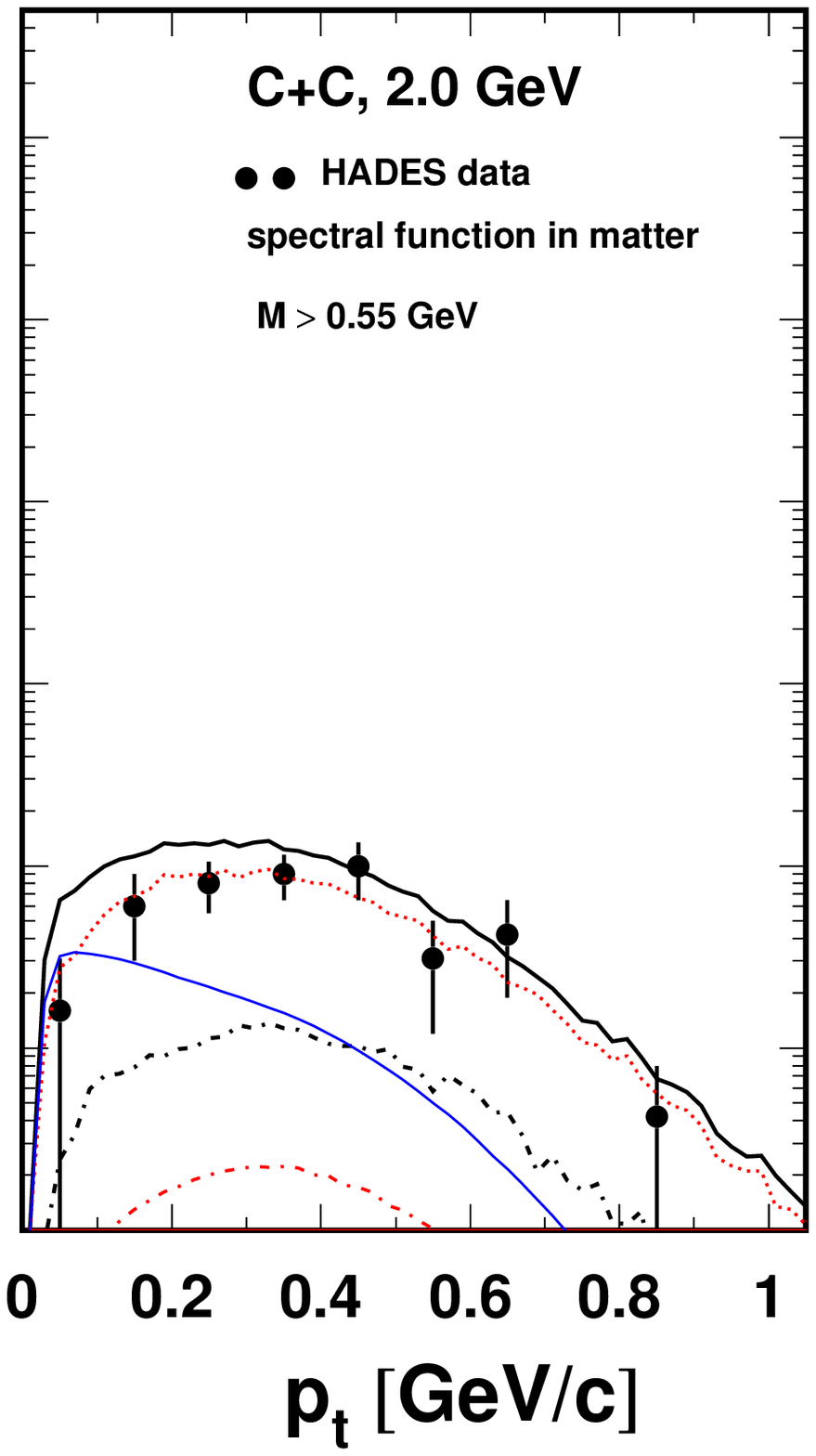,width=0.35\linewidth,angle=0} 
\vspace*{-1mm}
\caption{\it Transverse momentum spectra for three mass bins.
Data source: \cite{HADES_CC2_pt}. }
\label{fig_pt2_HADES}
\end{figure}

The transverse momentum spectra for three invariant mass bins
are exhibited in Fig.~\ref{fig_pt2_HADES}. One recognizes a good agreement 
with these multi-differential data.

\subsection{Effects in larger collision systems}

\begin{figure}[!htb]
\center
\epsfig{file=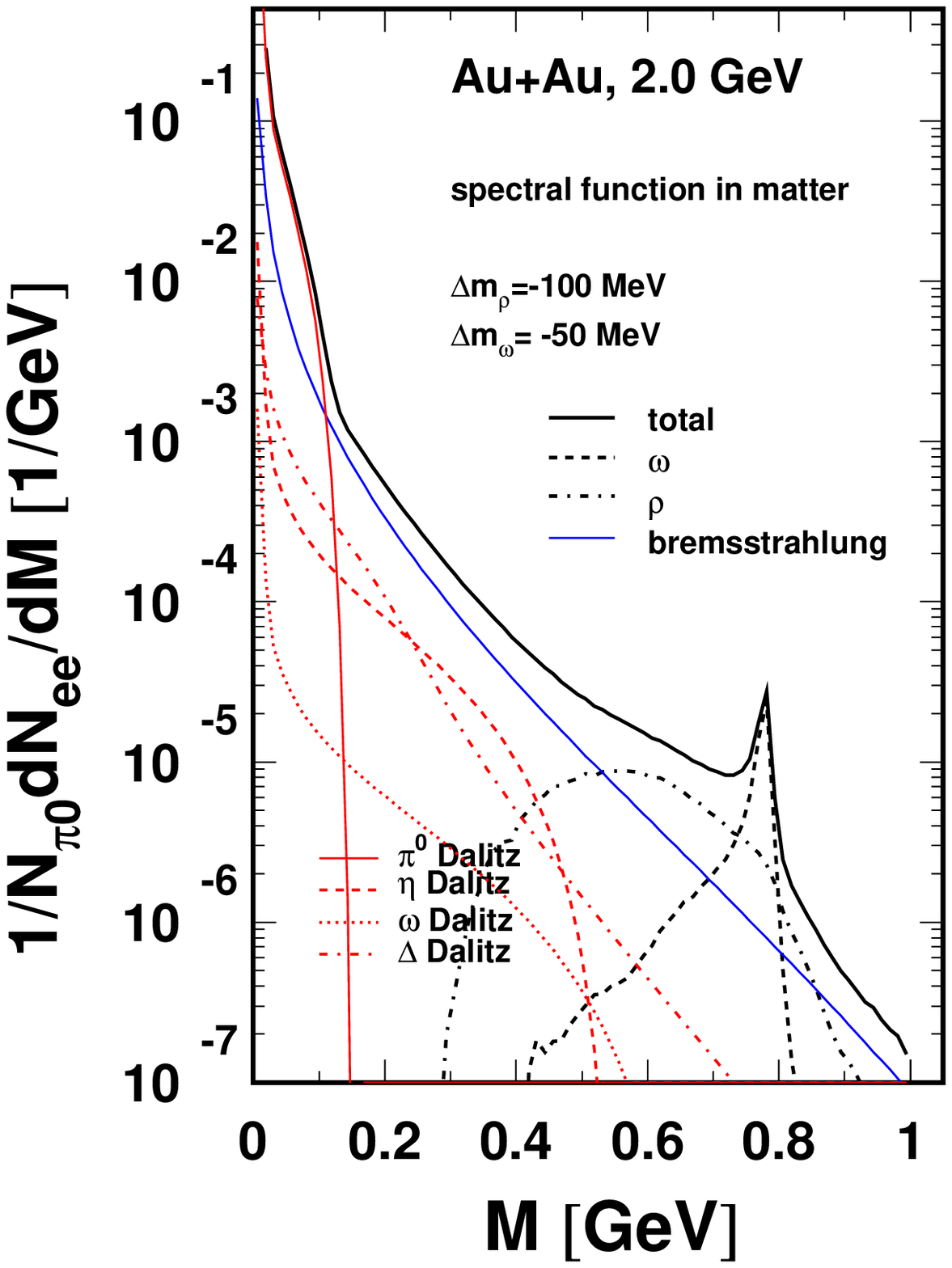,width=0.35\linewidth,angle=0} \hspace*{6mm}
\epsfig{file=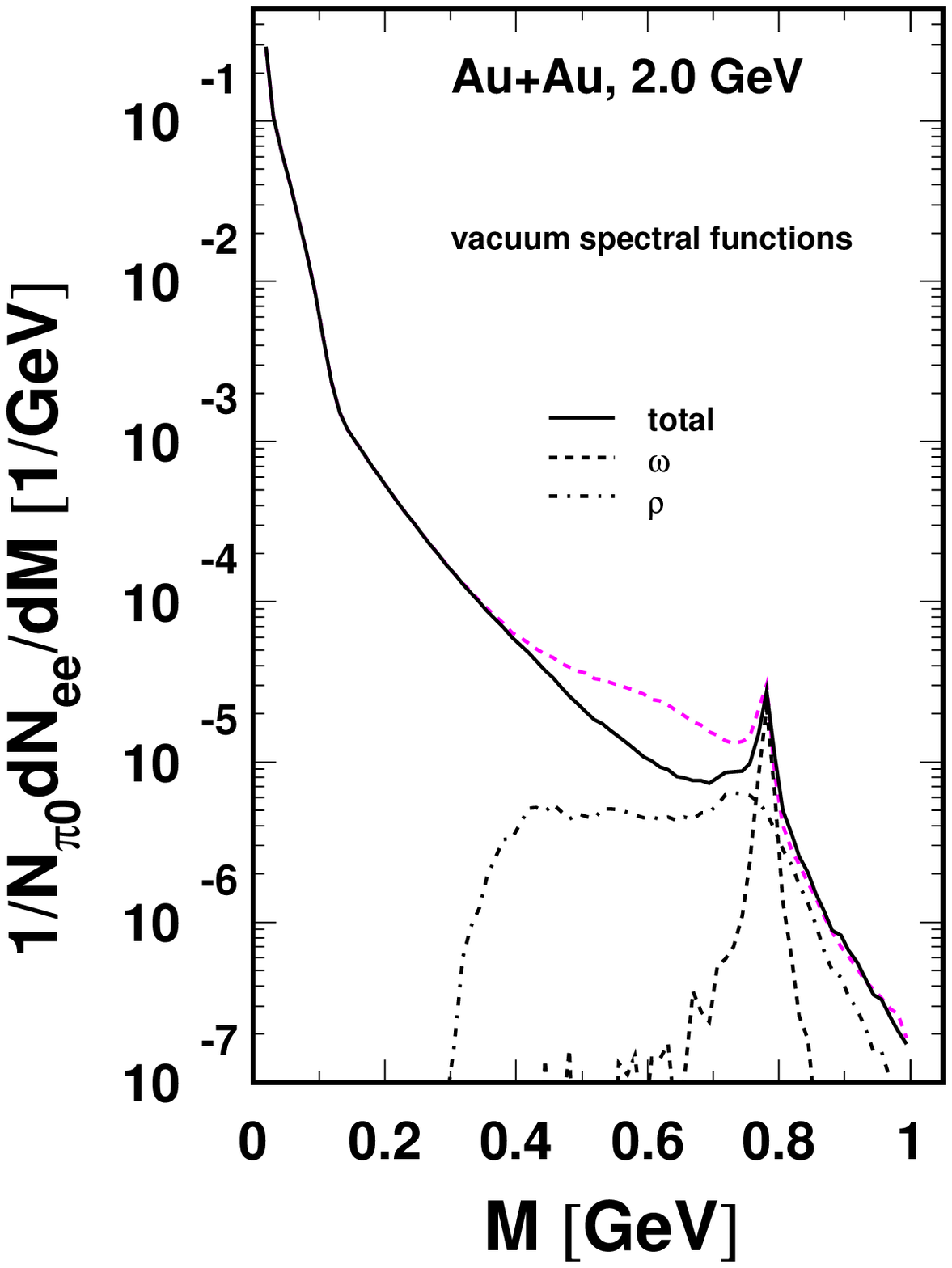,width=0.35\linewidth,angle=0}
\caption{\it Di-electron spectra for Au(2 AGeV) + Au. 
Left panel: In-medium spectral function is used and various contributions
to the total spectrum are shown.
Right panel: Vacuum spectral function (solid line) in comparison to the spectrum 
obtained with the in-medium spectral function (dashed violett line).}
\label{fig_HADES_Au}
\end{figure}

The C + C system is rather light and a consequence of it is that the maximum density
is about 2.5 $n_0$ (see Fig.~\ref{fig_test_ome}). A heavy system has a longer living 
high-density stage reaching densities of about 3.5 $n_0$. In Fig.~\ref{fig_HADES_Au}
we display the result of our calculations for a collision of Au + Au at
2 AGeV. Comparing the left and  the right hand part one recognizes
the larger effect of the assumed pole mass shifts of the mesons in contrast
to the C + C case shown in Figs.~\ref{fig_HADES_med} and \ref{fig_HADES_vac}.
Furthermore, it is clearly seen that the amount of di-electrons coming from 
$\rho$ mesons is relatively larger than for light systems. This can be understood
by the fact that during the longer collision time $\rho$ mesons can rapidly
decay and regenerate

\begin{figure}[!h]
\vskip -6mm
\center
\epsfig{file=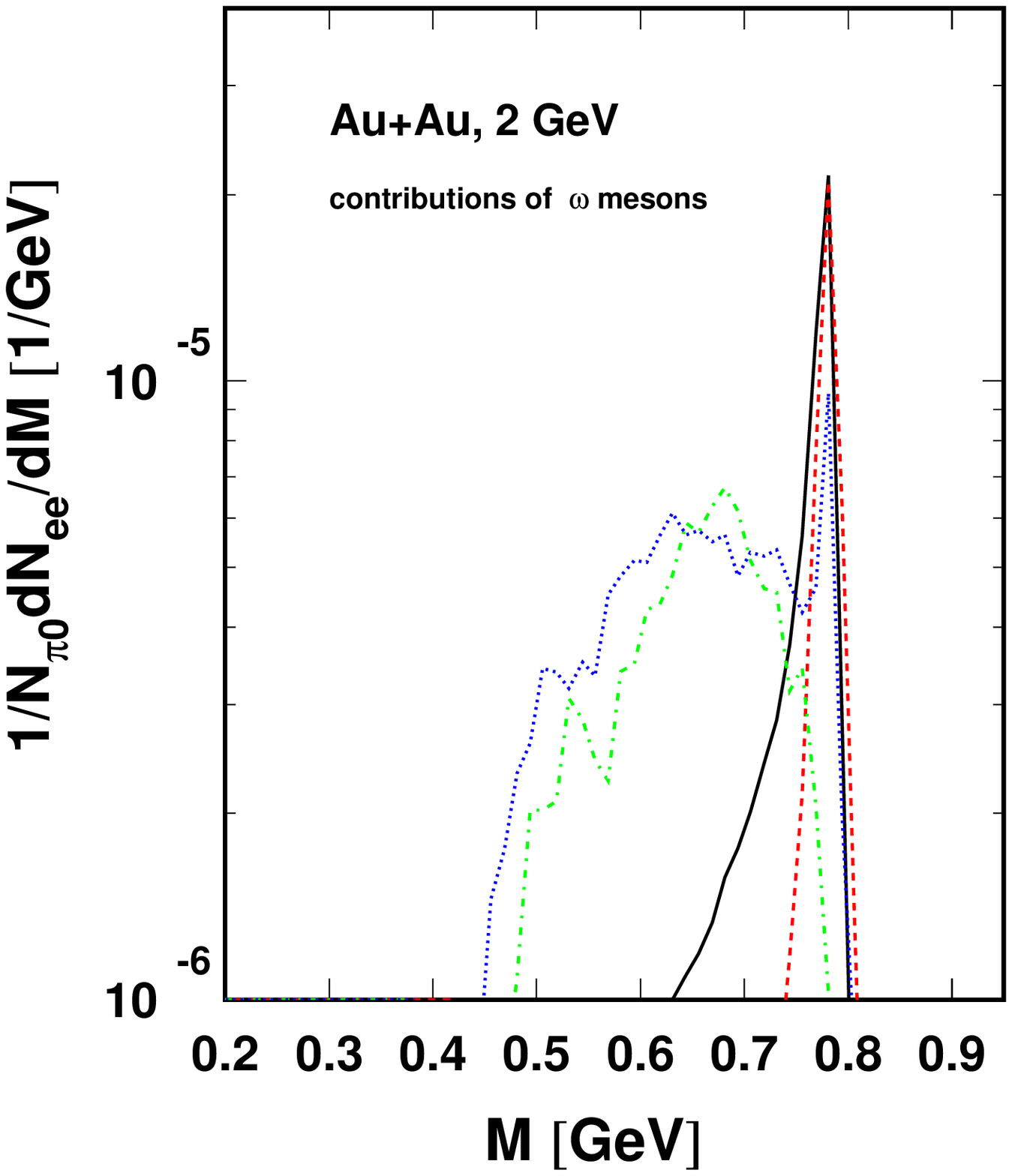,width=0.35\linewidth,angle=0} \hspace*{3mm}
\epsfig{file=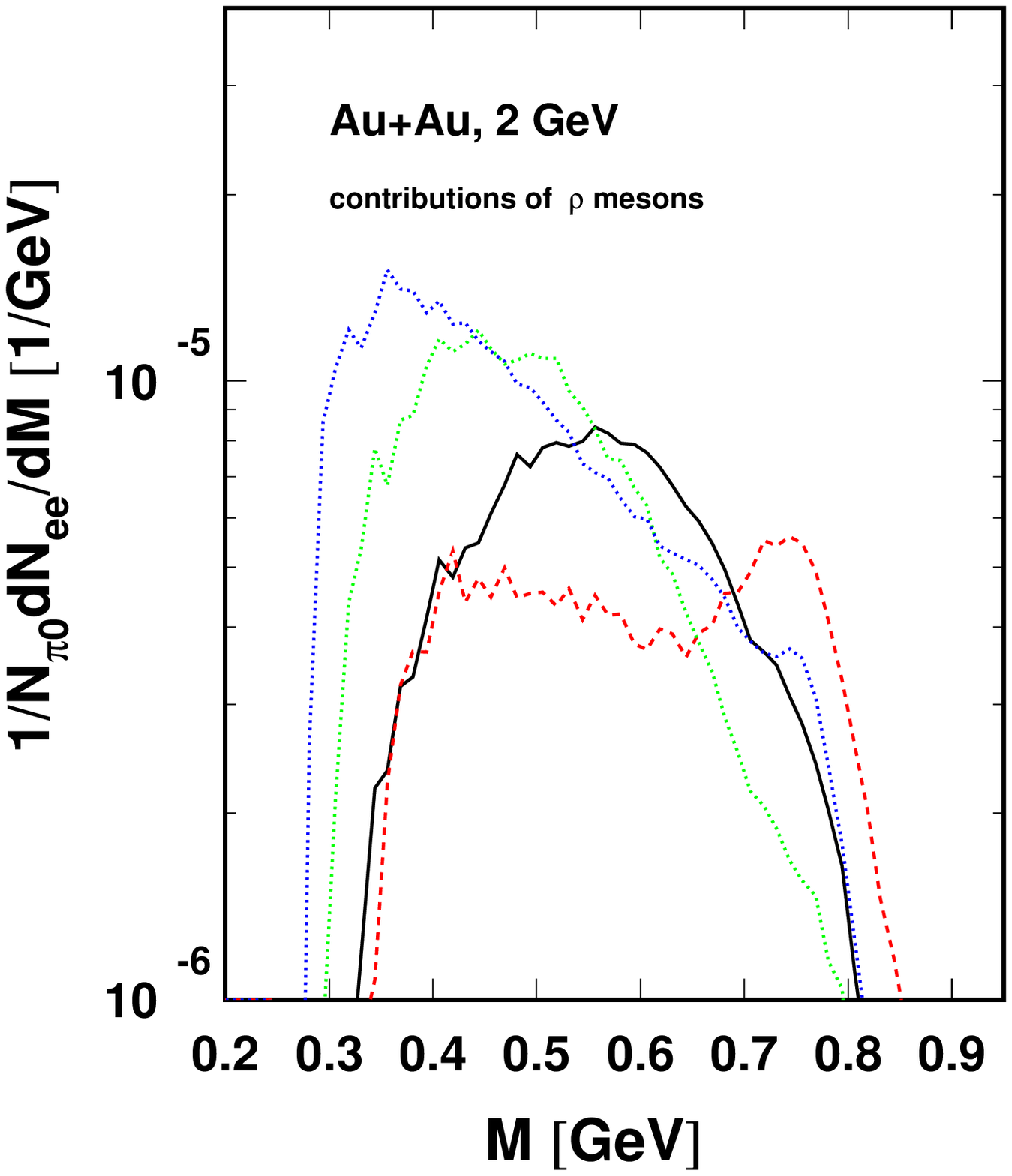,width=0.35\linewidth,angle=0}
\vspace*{-1mm}
\caption{\it The same as in Fig.~\ref{fig_comp_evol} but for central
Au + Au collisions.}
\label{fig_compAu_evol}
\end{figure}

The effects discussed with respect to Fig.~\ref{fig_comp_evol}
are more clearly seen for a larger collision system. 
Figure \ref{fig_compAu_evol} shows the dramatic change of the 
di-electron spectrum emitted from $\rho$ and $\omega$ mesons
in central collisions Au + Au.
The vacuum spectral function of the $\rho$ meson (right panel)
still shows the peak near the pole 
mass despite the $m_\rho^{-3}$ dependence. 
In case of medium modification the shape differs strongly from the vacuum one.

\begin{figure}[!htb]
\center
\vskip 3mm
\epsfig{file=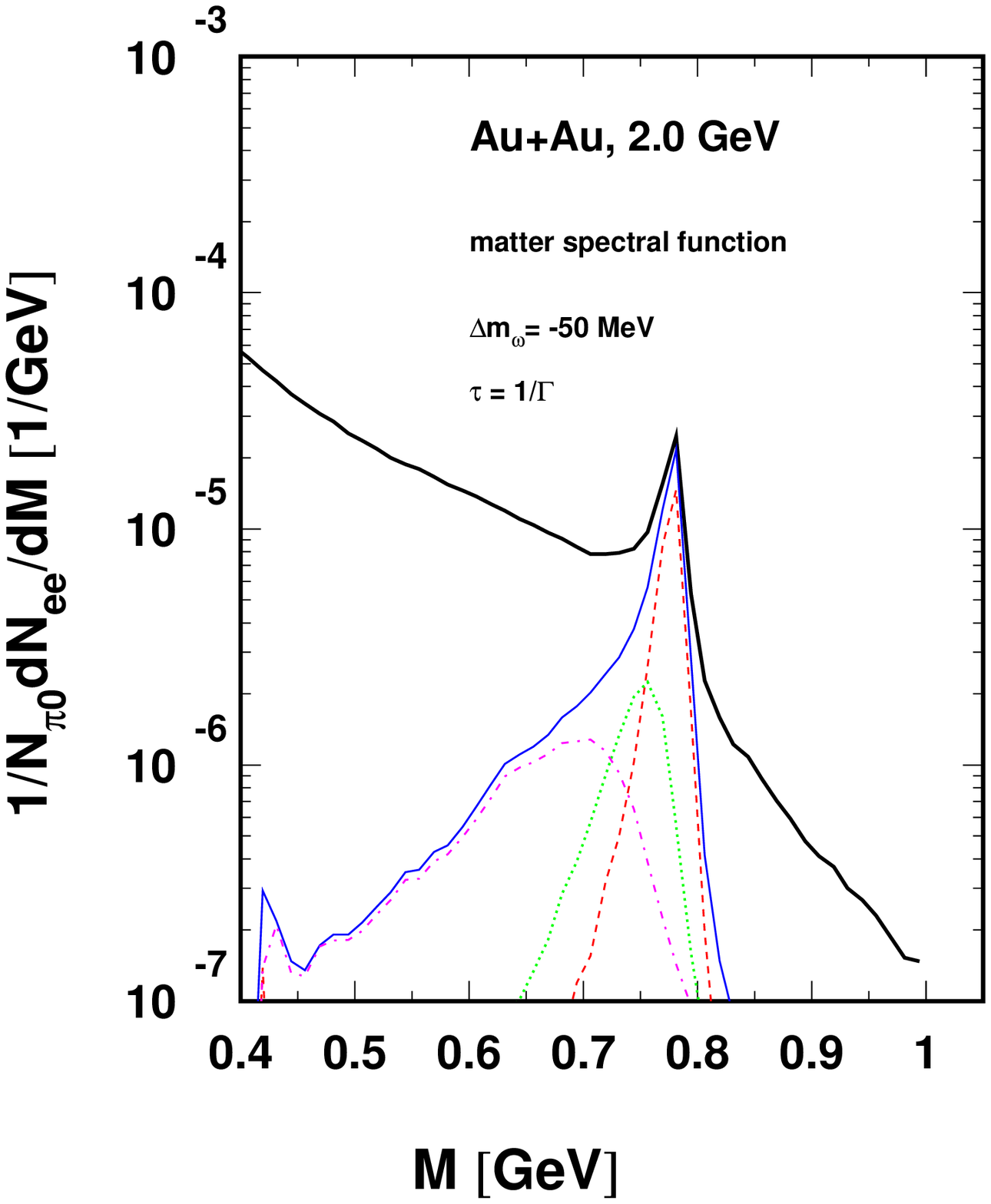,width=0.35\linewidth,angle=0} \hspace*{6mm}
\epsfig{file=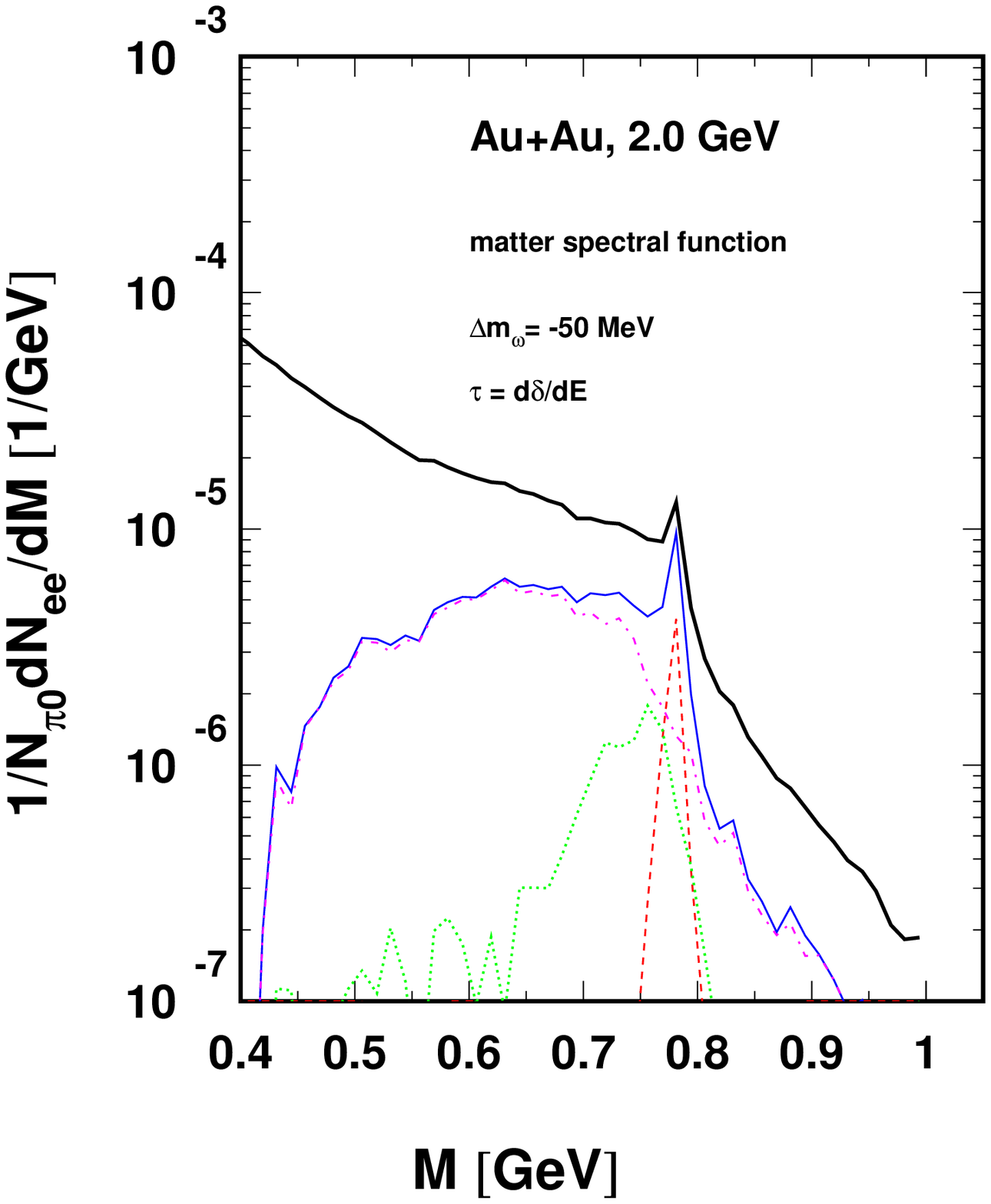,width=0.35\linewidth,angle=0}
\vspace*{-1mm}
\caption{\it
As in Fig.~\ref{fig_CC_dens} but for the heavy system $Au$+$Au$.}
\label{fig_Au_dens}
\end{figure}

Repeating the same analysis as in Fig.~\ref{fig_CC_dens} 
for central Au + Au collisions system we find that
di-electrons from the dense region (dot-dashed lines) have low masses
around 600 MeV and contribute roughly 10\% to the total $\omega$ 
yield (see Fig.~\ref{fig_Au_dens}). 
There is a remarkable difference between the outcome of the standard 
life time expression and Eq.~(\ref{decayrate}).

\section{Results at 1 AGeV}

\begin{figure}[!htb]
\center
\hspace*{-6mm}
\epsfig{file=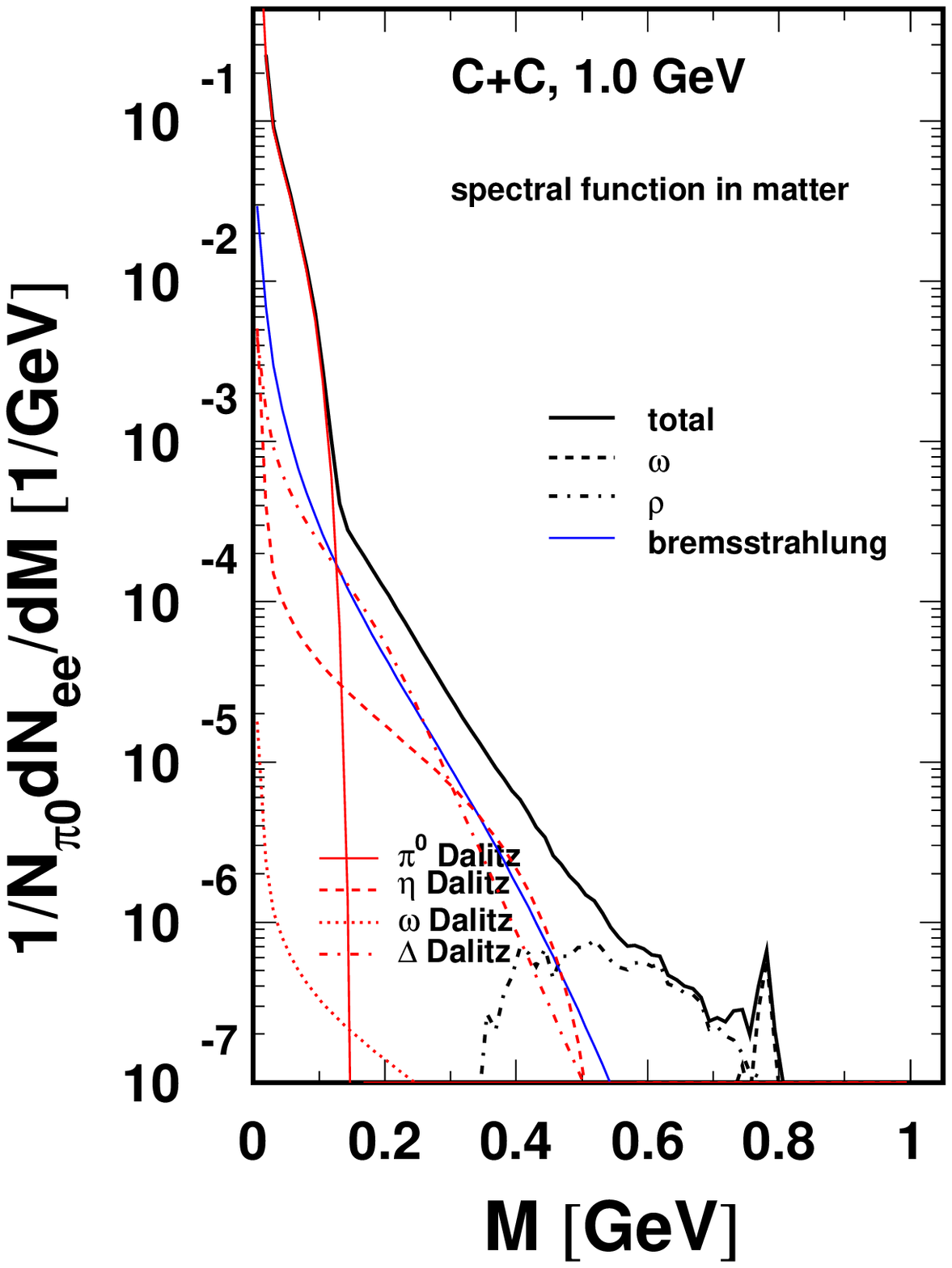,width=0.35\linewidth,angle=0} \hspace*{3mm}
\epsfig{file=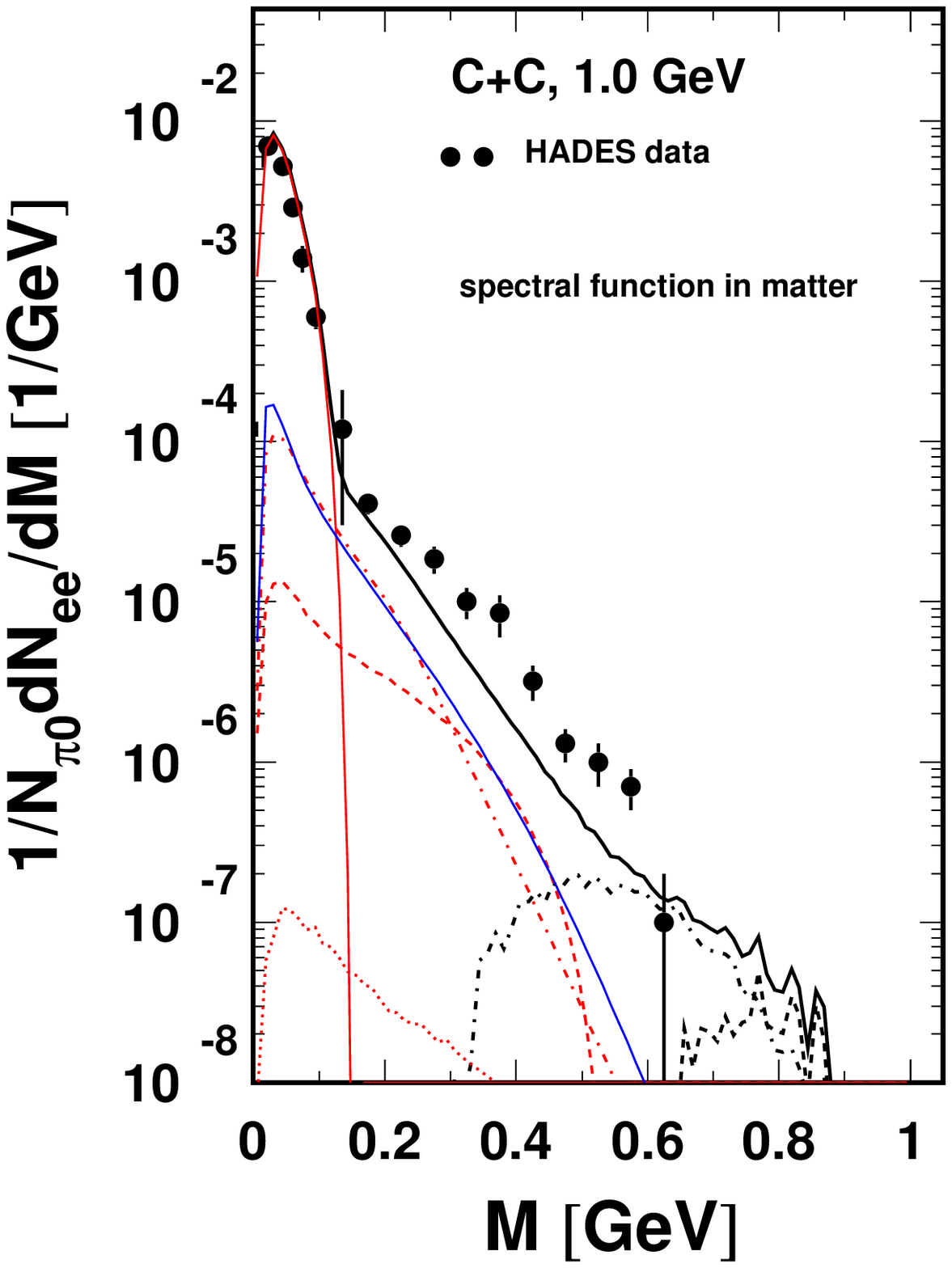,width=0.35\linewidth,angle=0}
\vspace*{-2mm}
\caption{\it Di-electron invariant mass spectrum
 for C(1 AGeV) + C calculated with in-medium spectral functions. 
 Left panel: Spectrum in full phase space. 
 Right panel: Comparison with HADES data \cite{HADES_1GeV}.
\label{fig_HADES_1m}}
\end{figure}

\begin{figure}[!htb]
\vspace*{-1mm}
\center
\hspace*{-1mm}
\epsfig{file=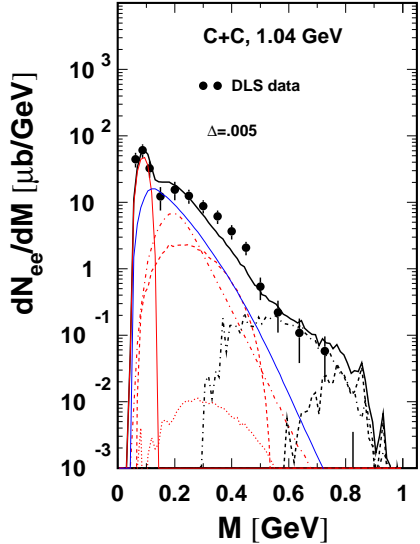,width=0.35\linewidth,angle=0}
\vspace*{-1mm}
\caption{\it Comparison of the di-electron spectrum 
(line codes as in Fig.~\ref{fig_HADES_1m})
calculated with the matter spectral functions with the 
the measurements of the DLS collaboration \cite{DLS}. The filter 
described on the DLS web page \cite{DLS_filter} was used.
\label{DLS_aata}}
\end{figure}

Recently new
HADES data are available also for C + C collisions at a bombarding
energy of 1 GeV per nucleon \cite{HADES_1GeV}. 
The excess energy is about 450 MeV,
thus only low energy tails of the $\rho$ and $\omega$ mesons play a role.
Nevertheless the $\rho$ mesons contribute essentially to the di-electron
spectrum above an invariant mass of 500 MeV since other sources are even
much smaller, see Figs.~\ref{fig_HADES_1m} and \ref{DLS_aata}.

We obtain a reasonable agreement with the measured HADES \cite{HADES_1GeV} 
(Fig.~\ref{fig_HADES_1m}) and DLS \cite{DLS} (Fig.~\ref{DLS_aata}) data,
but the data
at an invariant mass around 400 MeV are underestimated. The shoulder 
in the data could only be explained by a higher contribution of 
di-electrons coming from the Dalitz decay of the $\eta$ mesons. 
We calculate a production cross section  $\sigma_{\eta}=1.8$ mb which has
to be compared to the value $\sigma_{\pi^0}= 450$ mb. The experimental values
measured by \cite{TAPS}, $\sigma_{\eta}=1.5\pm0.4$ mb  and  
$\sigma_{\pi^0}= 287\pm21$ mb, give a ratio which is a factor of 1.3 larger
than our calculated value. Such an small increase of the $\eta$ yield 
could hardly improve the total di-electron spectra
and explain the shoulder at 400 MeV.

\begin{figure}[!htb]
\center
\vspace*{-0mm}
\hspace*{-1mm}
\epsfig{file=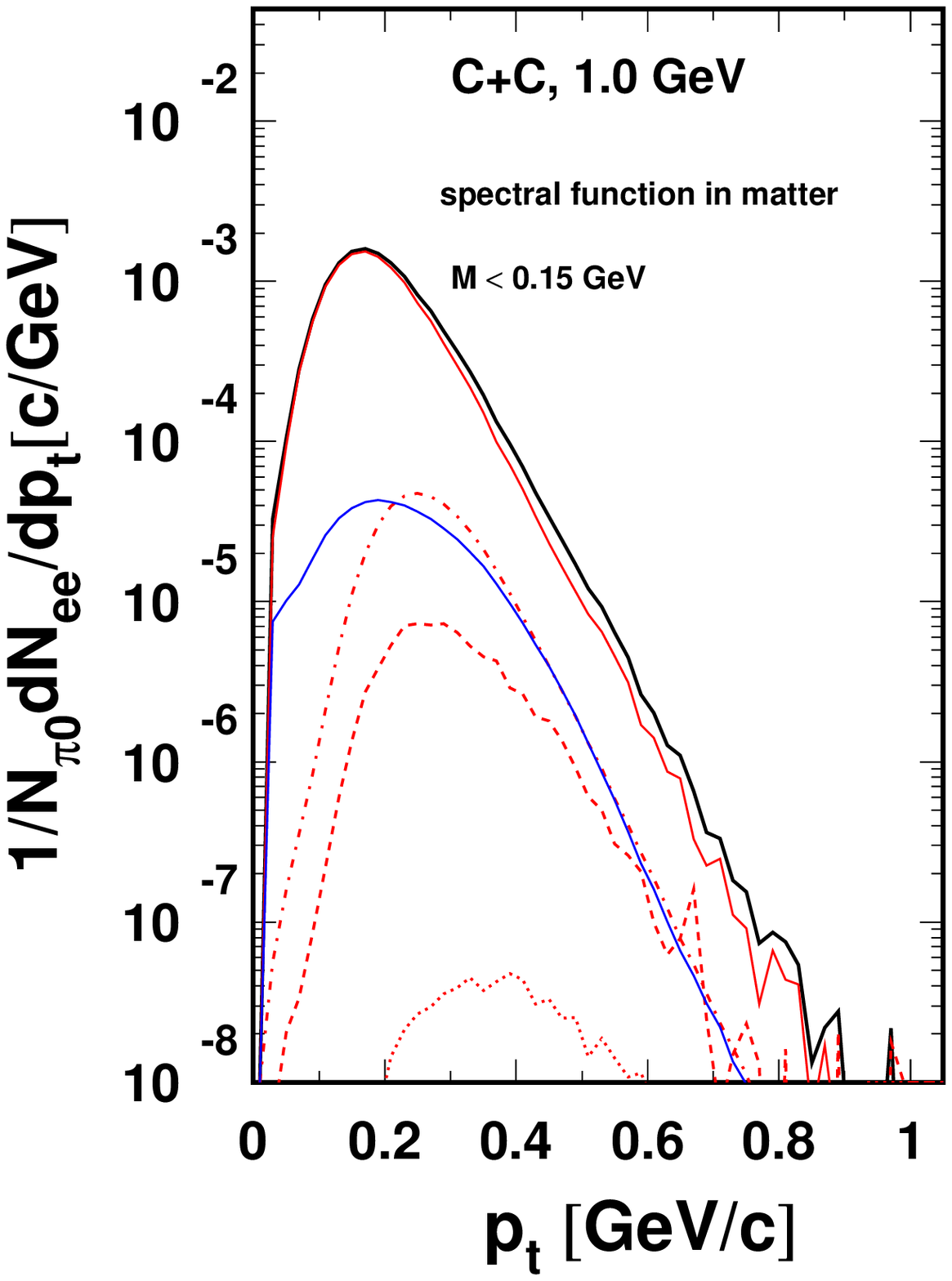,width=0.35\linewidth,angle=0} \hspace*{-17.3mm}
\epsfig{file=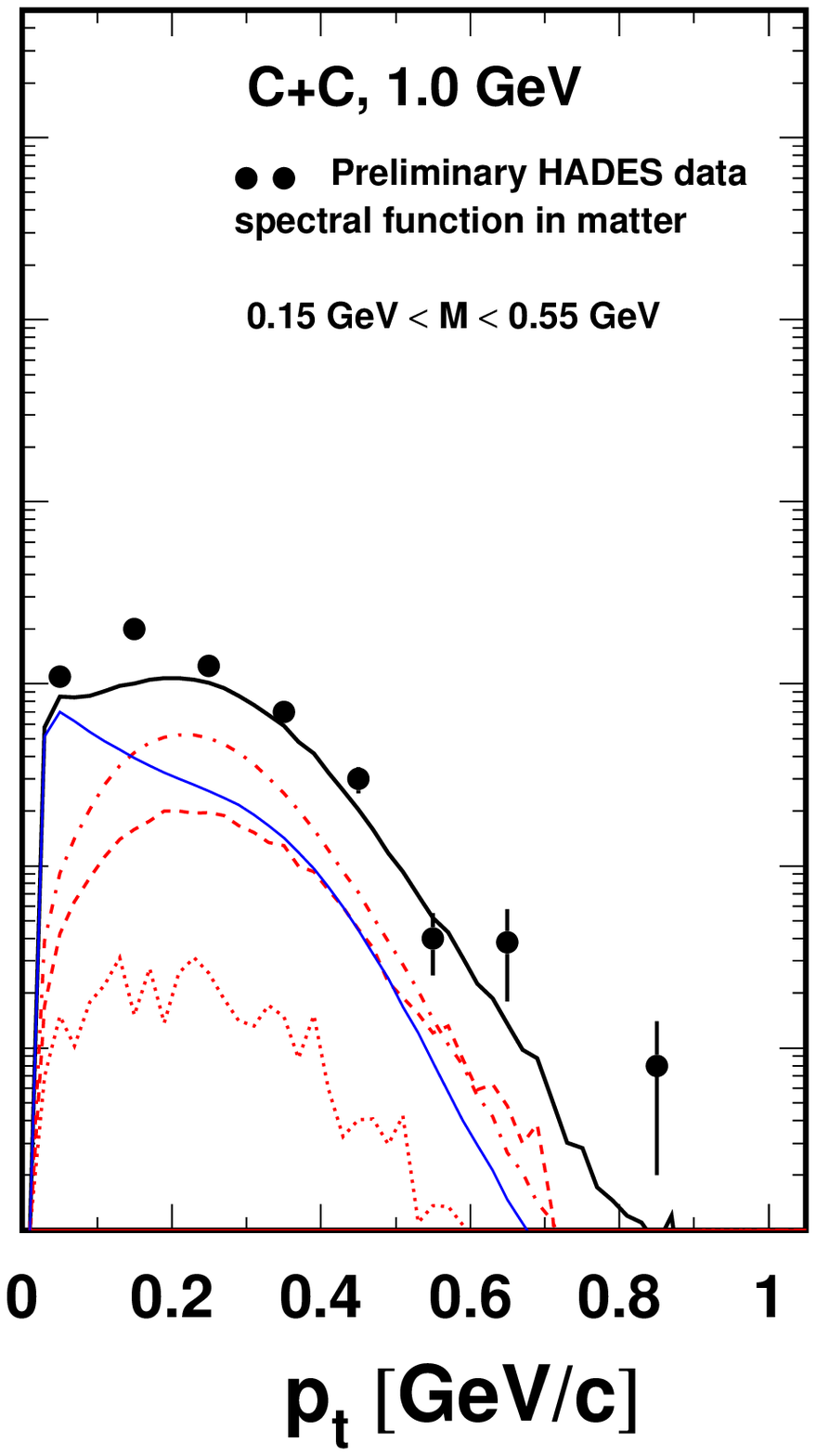,width=0.35\linewidth,angle=0} \hspace*{-17.3mm}
\epsfig{file=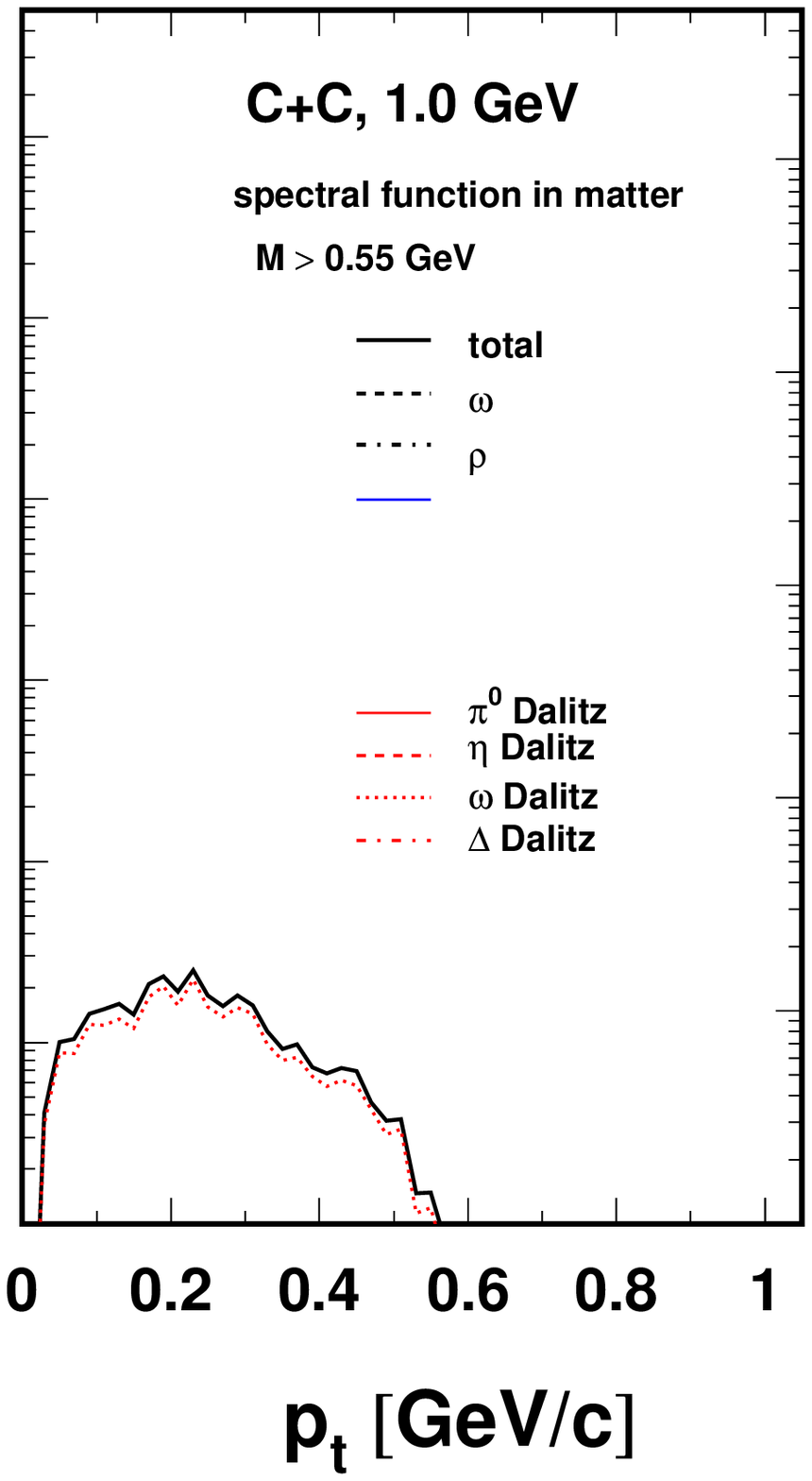,width=0.35\linewidth,angle=0}
\vspace*{-1mm}
\caption{\it Transverse momentum spectra for three mass bins at 1 AGeV bombarding energy.
Data source: \cite{Pachmayer}.}
\label{fig_pt1_HADES}
\end{figure}

Figure~\ref{fig_pt1_HADES} exhibits the transverse momentum spectra 
for three invariant mass
bins. A good agreement with available data can be stated. 

\begin{figure}[!htb]
\center
\vspace*{-3mm}
\epsfig{file=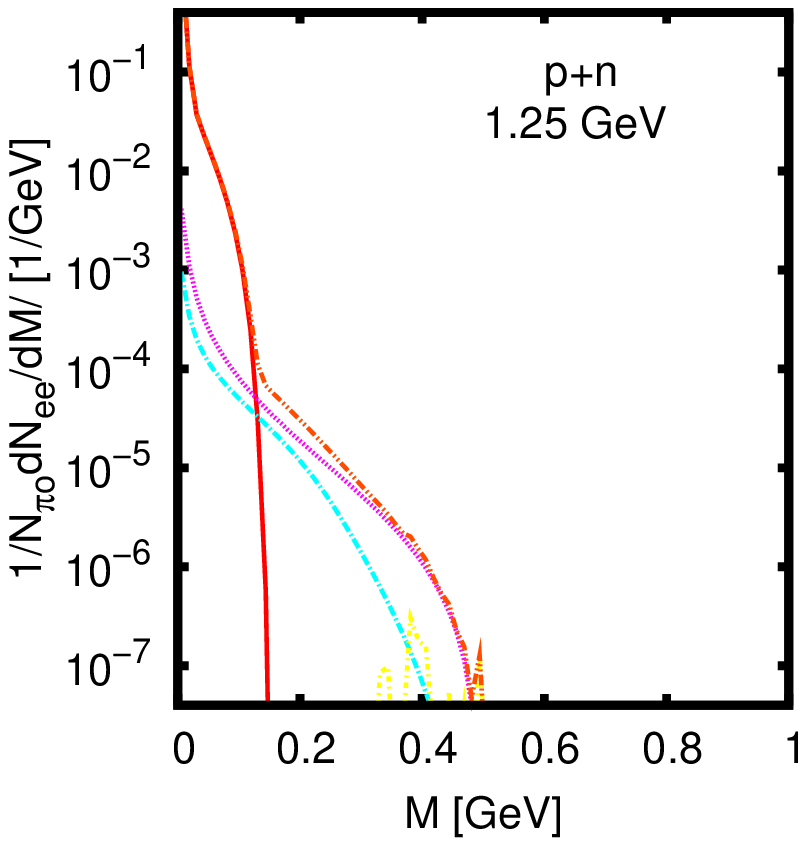,width=0.40\linewidth,angle=0} \hspace*{3mm}
\epsfig{file=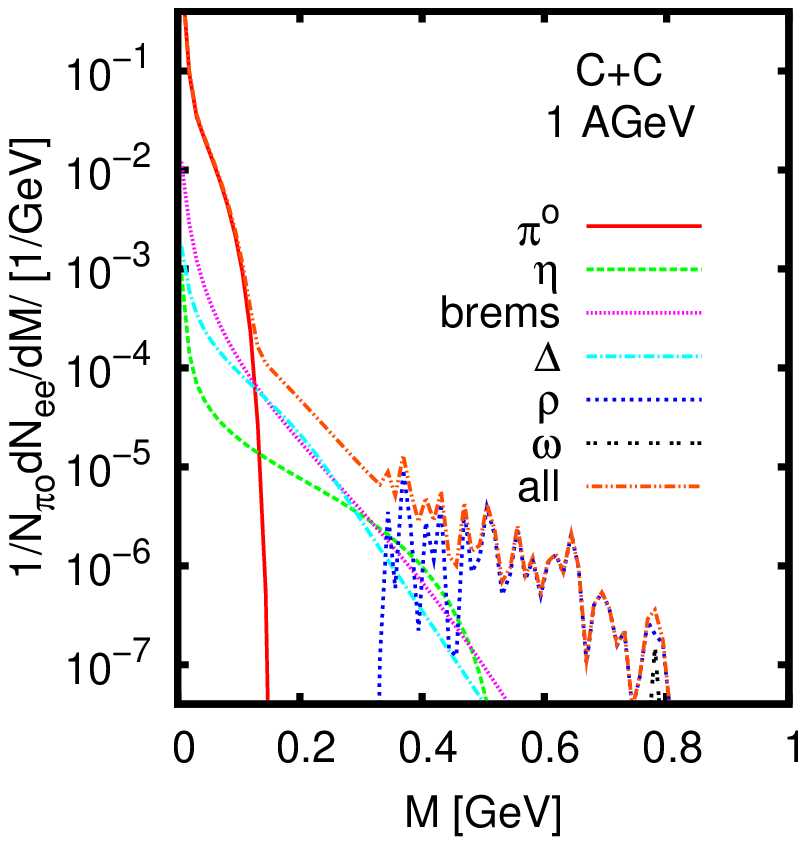,width=0.40\linewidth,angle=0}
 \caption{\it Comparison of invariant mass spectra for n + p at 1.25 GeV (left panel,
 line codes as in right panel)
 and C + C at 1 AGeV (right panel).}
\label{np_CC_1}
\end{figure}

Remarkable is the following scaling property. Comparing the normalized
di-electron spectra for the reaction n + p at 1.25 GeV and C + C at 1 GeV
(see Fig.~\ref{np_CC_1})
one recognizes that the spectra agree to a large extent despite of the
different systems and beam energies. 

\section{Summary}

In summary we have considered the propagation of broad resonances
within a kinetic theory (transport) approach to heavy-ion
collisions. Vector mesons are described by spectral functions
and these are evolved in space and time by a test-particle method.
The motivation for this work is a new generation of data on di-electrons.
The corresponding experiments are aimed at seeking for imprints
of chiral symmetry restoration as particular aspect of in-medium
modifications of hadrons. This lets us focus on the treatment
of $\rho$ and $\omega$ mesons. The wildly wide-spread predictions call
for an experimental clarification, but still heavy-ion data need
often the comparison with models to extract the wanted information
from data. 

We have utilized here the transport equations from Ref.~\cite{Cassing-Juchem00}
which are approximations of the much more involved Kadanoff-Baym equations 
\cite{CGreiner}.
Compared to an approach wherein the spectral function is frozen in after
creation the present framework let the spectral functions evolve towards
the vacuum spectral functions. Therefore, the in-medium modifications
are washed out, in particular, for the $\omega$ meson. 
In contrast to earlier expectations the $\omega$ peak does not suffer a significant
modification, even when assuming a strong hypothetical shift of the peak position.
Within the employed framework, medium modifications of $\rho$ and $\omega$
mesons are hardly seen in the di-electron spectra of small collision systems, 
even when using fairly strong 
and schematic assumptions for them. Only heavy collision systems seem to allow
still to identify the wanted medium modifications.
The elementary channels which contribute to the overall yields need better
control to arrive at firm conclusions on interesting many-body effects.

{\it Acknowledgements:} 
We gratefully acknowledge the continuous information by the
HADES collaboration, in particular R.\ Holzmann for delivering and assisting
us in using the acceptance filter routines. The work is supported by the German 
BMBF 06DR136, GSI-FE and the Hungarian OTKA T48833 and T71989.


\begin{thebibliography}{99} 
\bibitem{RW} 
R. Rapp, J. Wambach, Adv.\ Nucl.\ Phys.\  25, 1 (2000);\\
R. Rapp, H. van Hees, J. Wambach, arXiv:0901.3289 [hep-ph].
\bibitem{Itzak} 
I. Tserruya, arXiv:0903.0415. 
\bibitem{Hatsuda_Lee} 
T. Hatsuda, S.-H. Lee, Phys. Rev. C 46, 34 (1992).
\bibitem{Brown_Rho} 
G.E. Brown, M. Rho, Phys. Rept. 363, 85 (2002) 
and further references therein.
\bibitem{MML} 
S. Leupold, V. Metag, U. Mosel, arXiv:0907.2388. 
\bibitem{DLS} 
R.J. Porter et al. (DLS), Phys. Rev. Lett. 79, 1229 (1997);\\
W.K. Wilson et al. (DLS), Phys. Rev. C 57, 1865 (1998).
\bibitem{HADES} 
P. Salabura et al. (HADES), Nucl. Phys. A 749 (2005) 150.
\bibitem{HADES_PRL} 
G. Agakichiev et al. (HADES), Phys. Rev. Lett. 98, 052302 (2007).
\bibitem{HADES_1GeV} 
G. Agakichiev et al. (HADES), Phys. Lett. B 663 (2008) 43.
\bibitem{Rapp} 
R. Rapp, J. Wambach, Eur. Phys. J. A 6, 415 (1999). 
\bibitem{Gallmeister} 
B. K\"ampfer, K. Gallmeister, O.P. Pavlenko, C. Gale, Nucl. Phys. A 698, 424 (2002),  
Nucl. Phys. A 688, 939 (2001);\\
K. Gallmeister, B. K\"ampfer, O.P. Pavlenko, Phys. Rev. C 62, 057901 (2000);\\
K. Gallmeister, B. K\"ampfer, O.P. Pavlenko, Phys. Lett. B 473, 20 (2000). 
\bibitem{Kapusta_hydro} 
P. Huovinen, M. Belkacem, P. Ellis, J. Kapusta, Phys. Rev. C 66, 014903 (2002).
\bibitem{Wambach_Cassing} 
W. Cassing, E.L. Bratkovskaya, R. Rapp, J. Wambach, Phys. Rev. C 57, 916 (1998).
\bibitem{broadening_omega} 
Gy. Wolf, B. Friman, M. Soyer, Nucl. Phys. A 640, 129 (1998).
\bibitem{Knoll} 
J. Knoll, Prog. Part. Nucl. Phys. 42, 177 (1999).
\bibitem{our_first_attempt} 
H.W. Barz, B. K\"ampfer, Gy. Wolf, M. Zetenyi, e-Print: nucl-th/0605036.
\bibitem{Brat_Cass} 
E.L. Bratkovskaya, W. Cassing, Nucl. Phys. A 807, 214 (2008).
\bibitem{Aichelin} 
M. Thomere, C. Hartnack, Gy. Wolf, J. Aichelin, Phys. Rev. C 75, 064902 (2007).
\bibitem{Tuebingen} 
E. Santini, M.D. Cozma, A. Faessler, C. Fuchs, M.I. Krivoruchenko, B. Martemyanov,
Phys. Rev. C 78, 034910 (2008);\\ 
S. Vogel, H. Petersen, K. Schmidt, E. Santini, C. Sturm, J. Aichelin, M. Bleicher,
Phys. Rev. C 78, 044909 (2008);\\
M.D. Cozma, C. Fuchs, E. Santini, A. Faessler, 
Phys. Lett. B 640, 170 (2006).
\bibitem{Bleicher} 
K. Schmidt, E. Santini, S. Vogel, C. Sturm, M. Bleicher, H. St\"ocker, 
Phys. Rev. C 79, 064908 (2009).
\bibitem{wolf93} 
Gy. Wolf, W. Cassing, U. Mosel, Nucl. Phys. A 552, 549 (1993);\\
S. Teis, W. Cassing, M. Effenberger, A. Hombach, U. Mosel, Gy. Wolf, Z. Phys. A 356, 421 (1997).
\bibitem{method} G.F. Bertsch, S. Das Gupta, Phys. Rep. 160, 189 (1988).
\bibitem{Wolf90} 
Gy. Wolf, G. Batko, W. Cassing, U. Mosel, K. Niita, M. Sch\"afer, 
Nucl. Phys. A 517, 615 (1990).
\bibitem{barznaumann} 
H.W. Barz, L. Naumann, Phys. Rev. C68 041901 (2003);\\
H.W. Barz, M. Zetenyi, Gy. Wolf, B. K\"ampfer, Nucl. Phys. A 705, 223 (2002);\\
H.W. Barz, M. Zetenyi, Phys. Rev. C 69, 024605 (2004).
\bibitem{Kadanoff} 
L.P. Kadanofff and G. Baym, {\it Quantum statistical mechanics},
Benjamin, New York, 1962.
\bibitem{Cassing-Juchem00} 
W. Cassing, S. Juchem, Nucl. Phys. A 672, 417 (2000).
\bibitem{Leupold00} 
S. Leupold, Nucl. Phys. A 672, 475 (2000).
\bibitem{Eff_gammaA} 
M. Effenberger, E. L. Bratkovskaya, U. Mosel, Phys. Rev. C 60, 044614  (1999).
\bibitem{Eff_off-shell} 
M. Effenberger, U. Mosel, Phys. Rev. C 60, 051901 (1999).
\bibitem{Bratkovskaya} 
E.L. Bratkovskaya, Nucl. Phys. A 696, 761 (2001).
\bibitem{Lutz} 
M. Lutz, Gy. Wolf, B. Friman, Nucl  Phys. A 706, 431 (2002).
\bibitem{rho_models} 
W. Peters, M. Post, H. Lenske, S. Leupold, U. Mosel, Nucl. Phys. A 632, 109 (1998).
\bibitem{rho_models1} 
M. Post, S. Leupold, U. Mosel, Nucl. Phys. A 689, 753 (2001).
\bibitem{wolf97} 
Gy. Wolf, Heavy Ion Phys. 5, 281 (1997).
\bibitem{rho_models2} 
S. Leupold, W. Peters, U. Mosel, Nucl. Phys. A 628, 311 (1998).
\bibitem{Leupold01} 
S. Leupold, Nucl. Phys. A 695, 377 (2001).
\bibitem{baz} 
A.I. Baz, Ya.B. Zeldovich and A.M. Perelomov,
Rassejanie, reakcii i raspady v nerelativistikoi kvantovoi mekhanike,
Nauka, Moscow, 1971.
\bibitem{Danielewicz} 
P. Danielewicz, S. Pratt, Phys. Rev. C 53, 249 (1996).
\bibitem{SATURN} 
F. Hibou et al., Phys. Rev. Lett. 83, 492 (1999). 
\bibitem{COSY} 
S. Abd El-Samad et al., Phys. Lett. B 522, 16 (2001).
\bibitem{DISTO} 
F. Balestra et al., Phys. Rev. C 63, 024004 (2001). 
\bibitem{HERA} 
V. Flaminio et al., report CERN-HERA 84-10 (1984).
\bibitem{DISTOrho} 
F.  Balestra et al., Phys. Rev. Lett. 89, 092001 (2002).
\bibitem{Calen} 
H. Calen et al., Phys. Lett. B 366, 39 (1996).
\bibitem{Hibou} 
F. Hibou et al., Phys. Lett. B 438, 41 (1998).
\bibitem{Smyrski} 
J. Smyrski et al., Phys. Lett. B 474, 182 (2000).
\bibitem{Chiavassa} 
E. Chiavassa et al., Phys. Lett. B 322, 270 (1994). 
\bibitem{Kaptari_omega} 
L.P. Kaptari, B. K\"ampfer, Eur. Phys. J. A 23, 291 (2005). 
\bibitem{Kaptari} 
L. Kaptari, B. K\"ampfer, Nucl. Phys. A 764, 338 (2006).
\bibitem{PDG} 
D.E. Groom et al., Eur. Phys. J. C 15, 1 (2000).
\bibitem{Ernst} 
C. Ernst et al., Phys. Rev. C 58, 447  (1998).
\bibitem{pppaper} 
M. Z\'et\'enyi, Gy. Wolf, Phys. Rev. C 67, 044002 (2003).
\bibitem{Fuchs} 
M. I. Krivoruchenko, B. V. Martemyanov, A. Faessler, C. Fuchs, Ann. Phys. 296, 299 (2002).
\bibitem{Moniz} 
J.H. Koch, E.J. Moniz, N. Ohtsuka, Ann. Phys. 154, 99 (1984).
\bibitem{Trnka} 
D. Trnka et al. (CB-TAPS), Phys. Rev. Lett. 94, 192303 (2005).
\bibitem{Oset} 
M. Kaskulov, E. Hernandez, E. Oset, Eur. Phys. J. A 31, 245 (2007).
\bibitem{KEK} 
M. Naruki et al., Phys. Rev. Lett. 96, 092301 (2006).
\bibitem{Ellis} 
A.T. Martell, P.J. Ellis, Phys. Rev. C 69, 065206 (2004).
\bibitem{Leupold_omega} 
B. Steinmueller, S. Leupold, Nucl. Phys. A 778, 195 (2006);\\
P. Muehlich, V. Shklyar, S. Leupold, U. Mosel, M. Post, Nucl. Phys. A 780, 187 (2006).
\bibitem{QCDSR} 
R. Thomas, S. Zschocke, B. K\"ampfer, Phys. Rev. Lett. 95, 232301 (2005).
\bibitem{CLAS} 
M.H. Wood et al. (CLAS), Int. J. Mod. Phys. A 24, 309 (2009);\\
M.H. Wood et al. (CLAS), Phys. Rev. C 78, 015201 (2008);\\ 
R. Nasseripour et al. (CLAS), Phys. Rev. Lett. 99, 262302 (2007).
\bibitem{rho_models4} 
M. Harada, C. Sasaki, Phys. Rev. D 73, 036001 (2006).
\bibitem{rho_models5} 
J. Ruppert, T. Renk, B. M\"uller, Phys. Rev. C 73, 034907 (2006).
\bibitem{rho_models6} 
S. Leupold, Nucl. Phys. A 743, 283 (2004).
\bibitem{rho_models7} 
D. Cabrera, Nucl. Phys. A 721, 759 (2003).
\bibitem{Holzmann} 
R. Holzmann et al. (HADES), communiction on the HADES filter HAFT.
\bibitem{TAPS} 
R. Averbeck et al. (TAPS), Z. Phys. A 359, 65 (1997).
\bibitem{Zschocke} 
B. K\"ampfer, O.P. Pavlenko, S. Zschocke, Eur. Phys. J. A 17, 83 (2003).
\bibitem{HADES_CC2_pt} T. Eberl et al. (HADES),  Eur. Phys. J. C 49, 261 (2007);\\
M. Sudol, Ph.D. thesis, JWG University Frankfurt 2007.
\bibitem{DLS_filter} 
\begin{verbatim}
http://macdls.lbl.gov/DLS_WWW_Files/AA_Letter/AA.html.
\end{verbatim}
\bibitem{Pachmayer} Y. Pachmayer et al. (HADES),  J. Phys. G 35, 104159 (2008);\\
M. Sudol (HADES), Eur. Phys. J. C 62, 81 (2009);\\
Y. Pachmayer, Ph.D. thesis, JWG University Frankfurt 2008.
\bibitem{CGreiner} 
B. Schenke, C. Greiner, Phys. Rev. C 73, 034909 (2006). 
\end{thebibliography}
\end{document}